\documentclass[aps,prb,preprint,showpacs,amsfonts]{revtex4}

\usepackage{epsfig}
\usepackage{amsmath}
\usepackage{amssymb}

\newcommand{\beq}{\begin{equation}}
\newcommand{\eeq}{\end{equation}}
\newcommand{\beqarray}{\begin{eqnarray}}
\newcommand{\eeqarray}{\end{eqnarray}}
\newcommand{\sgn}[2][]{\ensuremath{\text{sgn}_{#1}(#2)}} 
\newcommand{\Kf}[1][\nu\sigma]{\ensuremath{\hat{F}^{}_{#1}}} 
\newcommand{\Kfdag}[1][\nu\sigma]{\ensuremath{\hat{F}^{\dagger}_{#1}}} 
\newcommand{\acom}[2]{\ensuremath{\left[#1,#2\right]_{+}}} 
\newcommand{\com}[2]{\ensuremath{\left[#1,#2\right]_{-}}} 
\newcommand{\half}{\ensuremath{\tfrac{1}{2}}}
\newcommand{\vF}{\ensuremath{v_{F}}} 
\newcommand{\kF}{\ensuremath{k_{F}}} 
\newcommand{\eF}{\ensuremath{e_{F}}} 
\newcommand{\Hc}{\ensuremath{\mbox{H.c.}}} 
\newcommand{\Ham}[1][]{\ensuremath{{\cal{H}}_{\text{\tiny{#1}}}}} 
\newcommand{\KD}[2]{\ensuremath{\delta_{#1,#2}}} 
\newcommand{\CO}[2][\alpha]{\ensuremath{\Lambda^{#2}_{#1}(k)}} 
\newcommand{\eq}[1]{Eq.~(\ref{#1})} 
\newcommand{\fig}[1]{Fig.~(\ref{#1})} 
\newcommand{\Sec}[1]{Sec.~\ref{#1}} 
\newcommand{\Ref}[1]{Ref.~\onlinecite{#1}} 

\begin{document}

\title{Charge order in the Falicov-Kimball model}
\author{P. M. R. Brydon and M. Gul\'{a}csi} 
\affiliation{ 
Department of Theoretical Physics, 
Institute of Advanced Studies, 
The Australian National University, Canberra, ACT 0200, Australia}

\date{\today}

\begin{abstract}
We examine the spinless one-dimensional Falicov-Kimball model (FKM)
below half-filling, addressing both the binary alloy and valence
transition interpretations of the model. Using a non-perturbative
technique, we derive an effective Hamiltonian for the occupation of
the localized orbitals, providing a comprehensive description of
charge order in the FKM. In particular, we uncover the
contradictory ordering roles of the forward-scattering and
backscattering itinerant electrons: the latter are responsible for the
crystalline phases, while the former produces the phase separation. We
find an Ising model describes the transition between the phase
separated state and the crystalline phases; for 
weak-coupling we present the critical line equation, finding
excellent agreement with numerical results. We consider several extensions of
the FKM that preserve the classical nature of the localized states. We also 
investigate a parallel between the FKM and the Kondo lattice model,
suggesting a close relationship based upon the similar orthogonality
catastrophe physics of the associated single-impurity models.
\end{abstract}

\maketitle

\section{Introduction} \label{sec:Intro}

The Falicov-Kimball Model (FKM) describes the interaction between
conduction electrons and localized atomic orbitals. The Hamiltonian of
the one-dimensional (1D) FKM for spinless Fermions is written
\beq
\Ham[FKM] = -t\sum_{j}\left\{c^{\dagger}_{j}c^{}_{j+1} +\Hc\right\}
+ \epsilon_{f}\sum_{j}n^{f}_{j} + G\sum_{j}n^{f}_{j}n^{c}_{j}
\label{eq:Intro:FKM} 
\eeq
where $t>0$ is the conduction ($c$) electron hopping, $\epsilon_{f}$
is the energy of the localized $f$-electron level, and $G$ is the
on-site interorbital Coulomb repulsion. 
The concentration of electrons is fixed at $n=
(1/N)\sum_{j}\left\{\langle{n^{f}_{j}}\rangle + 
\langle{n^{c}_{j}}\rangle\right\}$ where $N$ is the number of
sites. In this work we consider only the case $n<1$. We work
throughout at zero temperature $T=0$.

The FKM was originally developed as a minimal model of valence
transitions: continuous or discontinuous changes in the occupation of
the $f$ orbitals (the atomic ``valence'') were observed when varying
the coupling $G$ or the $f$-level energy
$\epsilon_{f}$.~\cite{FKMoriginal} Since only the distribution of
electrons across the two orbitals is of interest, the model has
traditionally been studied for spinless fermions. These early  
works, however, neglected an important feature of $\Ham[FKM]$: the
occupation of each $f$-orbital is a good quantum number and so may be
replaced in~\eq{eq:Intro:FKM} by its expectation value 
$n^{f}_{j}\rightarrow\langle{n^{f}_{j}}\rangle=0,1$. 
It was quickly realized that in many physical systems displaying a
valence instability (e.g. SmB$_{6}$ and Ce), this is an
inappropriate idealization. Instead of a mixture of atoms with
different integer valence, in these materials each atomic orbital
exists in a superposition of its different occupancy
states.~\cite{LRP81} Although the FKM was modified to include this
quantum behaviour by the addition of a $c$-$f$ hybridization
term,~\cite{Vfinite,Liu&Ho} it has now been superseded 
as a model of valence transitions by the periodic Anderson
model.~\cite{C86}  

The FKM was reinvented by Kennedy and Lieb in 1986 as a simple model
of a binary alloy.~\cite{KL86} Assuming fixed $c$- and $f$-electron
populations, the sites with occupied and unoccupied $f$ orbitals may
be regarded as 
different atomic species A and B respectively. The Coulomb repulsion
$G$ is interpreted as the difference between the single-particle
energies of the two atoms. For this so-called crystallization problem
(CP), the ground state is defined as the configuration of the two
atomic species ($f$ electrons) that minimizes the energy of the
$c$ electrons. The ordering of the different atomic constituents in a
binary alloy is an important theoretical and experimental problem: in
realistic systems a large range of ordered structures are observed,
although the electronic mechanisms responsible for these phases have
remained largely obscure.~\cite{deFontaine,D:OPSA} By studying a simple 
model such as the FKM, some insight into the origin
of the charge order might be obtained.

Kennedy and Lieb analyzed~\eq{eq:Intro:FKM} for a bipartite lattice
at half-filling and equal concentrations of $c$ and
$f$ electrons. In the limit of $T=0$ and strong-coupling,
they proved that the $f$ electrons occupied one sublattice only, the
so-called checkerboard state. This crystalline state is,
however, unique to half-filling: for all other fillings, the 
$G\rightarrow\infty$ ground state is the so-called segregated (SEG)
phase.~\cite{seg1D,segND} The SEG phase is characterized by the
$f$ electrons forming a single cluster, arranged in such a manner as to
present the smallest perimeter with the rest of the lattice, which is
occupied by the $c$ electrons.
These strong-coupling results hold for all 
dimensions $d$. At weak- and intermediate-coupling, the situation is
considerably more complicated: for $d=1$, both
analytic~\cite{GLM93,FGM96} and numeric~\cite{FF90,GUJ94,GJL96} 
studies have revealed a myriad of different crystalline orderings of
the $f$ electrons. The SEG phase is also realized, but not as
ubiquitously as at strong-coupling.
Intriguingly, for certain $c$- and $f$-electron
fillings, the system is unstable towards a special phase-separated
state, where the ground state configuration of the localized electrons
is a mixture of a crystalline phase and the state with completely
empty or full localized orbitals.~\cite{FB93,FGM96,GJL96} Work in higher
dimensions has revealed similar behaviour;~\cite{2D} the understanding
of the $d\rightarrow\infty$ limit phase diagram is particularly
advanced.~\cite{DMFT}  

Contemporary with Kennedy and Lieb's work, Brandt and Schmidt
introduced the FKM as an exactly-solvable model of a ``classical''
valence transition.~\cite{BS} The distribution of the electron weight between
the two orbitals is not fixed, but instead determined by the interations. 
Quantum effects such as superposition of orbital states are ignored: as in the
CP, the valence transition problem (VTP) is also concerned with the
configuration adopted by the available $f$ electrons. Despite the similarity
between the two interpretations, 
apart from the $d\rightarrow\infty$ limit~\cite{DMFT} and the $d=1$
half-filling case,~\cite{F95a,F95b} very little is known about the
ground states of the VTP. Since the ordered configurations found for
the CP occur over a 
wide range of different fillings and coupling strengths, we can
nevertheless expect that the VTP has a similarly rich phase diagram. 

Although an impressive catalog of charge-ordered phases has been
assembled for the 1D FKM, only the weak-coupling crystalline phases
are easily explicable as due to the $c$-electron backscattering off
the localized orbitals. The mechanism responsible for the
weak-coupling segregated and phase separated states remains unknown;
the competition between 
crystallization and the segregation is also poorly comprehended. In
this paper, we expand upon our previous work,~\cite{BG06} outlining a
comprehensive theory of the charge order in the FKM, with
particular emphasis on the phase separation.  

To describe the $c$ electrons, we use the well-known non-perturbative
bosonization technique, 
specially adapted to account for the presence of localized orbitals. 
We then canonically transform the bosonized FKM, rewriting the
Hamiltonian in a new basis that reveals the origin of the
phase separation to be the $c$-electron delocalization. Such a
mechanism has previously been proposed to account for the
ferromagnetic phase in the 1D Kondo lattice model (KLM),~\cite{HG}
pointing to a nontrivial connection between the two models based upon
orthogonality-catastrophe physics.
After simple manipulation of the transformed Hamiltonian we decouple
entirely the $c$ and the $f$ electrons, obtaining an Ising-like
effective Hamiltonian describing only the occupation of the
$f$ orbitals. The competition between the segregation and
crystallization is clearly evident in this effective model:
at weak-coupling we find the backscattering crystallization dominates
the physics; with increasing $G$, however, the electron delocalization
drives the system into the SEG phase. We verify that both
crystallization and segregation are present also in the VTP.

Our paper is arranged as follows: in~\Sec{sec:B} we give a brief
outline of our bosonization procedure, and present $\Ham[FKM]$ in the
bosonic form. We proceed to a description of the canonical transform
in~\Sec{sec:CT}, including a discussion of the resulting terms. 
We argue in~\Sec{sec:EH} for the derivation of the effective
Hamiltonian 
for the localized $f$ orbitals from the canonically-transformed
Hamiltonian; this is subsequently used in~\Sec{sec:PD} to interpret
the numerically-determined phase diagrams for the CP (\Sec{subsec:CP})
and the VTP (\Sec{subsec:VTP}).
We also present a brief analysis of several extensions of the FKM
in~\Sec{sec:EFKM}, specifically intraorbital nearest-neighbour
interactions and the introduction of spin, focusing upon 
possible alteration of the CP phase diagram. We conclude
in~\Sec{sec:conclusions} with a summary of our results and the outlook
for further work. 

\section{Bosonization} \label{sec:B}

The technique of bosonization has for many years been used to study
the critical properties of one-dimensional many-electron
systems.~\cite{G:QP1D} It relies upon the remarkable fact that
an effective low-energy description of such systems may be constructed
in terms of bosonic fields: this representation is usually much easier
to manipulate than the equivalent fermionic form. 
The bosonization of a tight-binding Hamiltonian is often performed in the
continuum limit where the lattice spacing 
$a\rightarrow0$;~\cite{G:QP1D} 
this approach is, however, inappropriate for systems involving
localized electron states.
For the itinerant electrons in the FKM, however, the usual
bosonization approach can be generalized to account for the
presence of the localized $f$ electrons. As explained
in~\Ref{G04}, this is accomplished by imposing a finite cut-off 
$\alpha>a$ on the wavelength of the bosonic density
fluctuations. Below we summarise our methodology.

The Bose representation is most conveniently written in terms of the
dual Bose fields. For a system of length $L\gg{a}$ we have
\beqarray
\phi(x_{j}) &=&
-i\sum_{\nu}\sum_{k\neq0}\frac{\pi}{kL}\rho_{\nu}(k)\CO{}{e}^{ikx_j}
\label{eq:B:phi}\\
\theta(x_{j}) &=&
i\sum_{\nu}\sum_{k\neq0}\nu\frac{\pi}{kL}\rho_{\nu}(k)\CO{}{e}^{ikx_j} \label{eq:B:theta}
\eeqarray
At the core of the bosonization technique are the chiral density
operators
\beq
\rho_{\nu}(k) =
q\sum_{0<\nu{k^{\prime}}<\pi/a}c^{\dagger}_{k^\prime-k}c_{k^{\prime}} 
\label{eq:B:rho} 
\eeq
which describe coherent particle-hole excitations about the right and
left Fermi points: as subscript (otherwise) we have $\nu=R(+)$, $L(-)$
respectively. The $\rho_{\nu}(k)$ are the basic bosonic objects,
obeying the standard commutation relations  
\beq
\com{\rho_{\nu}(k)}{\rho_{\nu'}(k')} =
\KD{\nu}{\nu'}\KD{k}{-k'}\frac{\nu{kL}}{2\pi}
\label{eq:B:comdo} 
\eeq
for wave vectors $|k|<\frac{\pi}{\alpha}$. The physical significance
of the Bose fields is as potentials: $\partial_{x}\phi(x_j)$ and
$\partial_{x}\theta(x_j)$ are respectively proportional to
the departure from the noninteracting values of the average electron
density and current at $x_{j}$.

The bosonic wavelength cut-off is enforced in \eq{eq:B:phi} and
\eq{eq:B:theta} by the function $\CO{}$ which has the approximate form   
\beq
\CO{} \approx \left\{
\begin{array}{cl}
1 & |k|< \frac{\pi}{\alpha} \\
0 & \text{otherwise}
\end{array} \right. \label{eq:B:CO}
\eeq
We expect that $\CO{}$ is a smoothly varying function of $|k|$,
reflecting the gradual change in the nature of the density
fluctuations. We require, however, that $\CO{}$ be not too 
`soft', i.e. $\CO{m}\approx\CO{}$ for $m=2,3,4$. The cut-off
essentially `smears' the Bose fields over the length $\alpha$ below
which the density operators do not display bosonic characteristics.
The commutators of the Bose fields reflect this smearing, with
important consequences for our analysis:
\beqarray
\com{\phi(x_{j})}{\theta(x_{j'})} &=&
\frac{i\pi}{2}{\sgn[\alpha]{x_{j'}-x_{j}}} \label{eq:B:comphitheta}\\
\com{\partial_{x}\phi(x_{j})}{\theta(x_{j'})} &=&
-i\pi\delta_{\alpha}(x_{j'}-x_{j}) \label{eq:B:comdelphitheta}
\eeqarray
$\sgn[\alpha]{x}$ and $\delta_{\alpha}(x)$ are the $\alpha$-smeared
sign and Dirac delta functions respectively. The precise
form of these functions depends upon $\CO{}$ [see~\Sec{subsec:SEGint}].

As is customary, we linearize the $c$-electron dispersion about the two
Fermi points. This allows a decomposition of the $j$-site annihilation
operator in terms of states in the vicinity of $\kF$ (the right-moving
fields) and $-\kF$ (the left-moving fields):
$$
c_{j} \approx c_{Rj}e^{i\kF{x_j}} + c_{Lj}e^{-i\kF{x_j}}
$$
Remarkably, the density operators $\rho_{\nu}(k)$ generate the entire
state space of the linearized Fermion Hamiltonian. A Bose
representation for the $c_{\nu{j}}$ may then be derived
by requiring that it correctly reproduces the Fermion
anticommutators and noninteracting expectation values. This leads to
the fundamental bosonization identity
\beq
c_{\nu{j}} =
\sqrt{\frac{{A}a}{\alpha}}\Kf[\nu]\exp\left(-i\nu\left[\phi(x_j) -
\nu\theta(x_j)\right]\right) \label{eq:B:b_identity}
\eeq
We note that this identity is only rigorously true in the
long-wavelength limit; \eq{eq:B:b_identity} may not correctly
reproduce the short-range ($<{\cal{O}}(\alpha)$) properties of the
$c_{\nu{j}}$. 
The dimensionless parameter $A$ is a normalization constant dependent
upon the cut-off function. The Klein factors $\Kf[\nu]$ obey the simple
algebra:   
\beq
\acom{\Kf[\nu]}{\Kf[\nu']} = 2\Kf[\nu]\Kf[\nu]\KD{\nu}{\nu'}, \qquad
\acom{\Kf[\nu]}{\Kfdag[\nu']} = 2\KD{\nu}{\nu'}
\eeq
The Klein factors act as ``ladder operators'': since the Bose fields only
operate within subspaces of constant particle number we require
operators to move between these different subspaces if we are to
regard equation \eq{eq:B:b_identity} as an operator identity. That is,
$\Kf[\nu]$ may be thought of as lowering the total number of
$\nu$-moving electrons by one. 

Using standard field-theory methods, the bosonization identity
\eq{eq:B:b_identity} may be used to derive the Bose representation for
any string of Fermion operators. Of particular note is the
representation for the $j$-site occupancy operator, $n^{c}_{j}$:
\beqarray
n^{c}_{j} & \approx &
\sum_{\nu,\nu^{\prime}}c^{\dagger}_{\nu{j}}c^{}_{{\nu^\prime}j}e^{-i(\nu-\nu')\kF{x_j}}
\notag \\
&=& n^{c}_{0}-\frac{a}{\pi}\partial_{x}\phi(x_j) +
\frac{Aa}{\alpha}\sum_{\nu}\Kfdag[\nu]\Kf[-\nu]e^{i2\nu\phi(x_j)}e^{-i2\nu{k}_{F}x_{j}} \label{eq:B:b_nc}
\eeqarray
The first term on the RHS, $n^{c}_{0}$, is the noninteracting
$c$-electron concentration; the second term gives the departure from
this value in the interacting system and is due entirely to forward
scattering ($\nu\rightarrow\nu$ processes); the third term is the first order
backscattering ($\nu\rightarrow-\nu$ processes) correction. Higher order
backscattering corrections are neglected. 

\subsection{The Hamiltonian in Boson Form} \label{subsec:BHam}

We bosonize the FKM Hamiltonian using the above methodology.
Since only the itinerant $c$ electrons can be bosonized, we
require that there be a finite population in the noninteracting
$c$-electron band. For the CP this simply requires
us to assume finite concentrations of the two species, $n^{c}$ and
$n^{f}$ for the $c$ and $f$ electrons respectively, which do
not change with the addition of the interaction term. For the VTP,
we impose the condition that the $f$-level coincides with the
Fermi energy in the noninteracting system: as we consider only the
case $n<1$, we limit ourselves to
$-2t<\epsilon_{f}<-2t\cos(\pi{n}/a)$. For $\epsilon_{f}$ outside this
range, our bosonization approach does not work. We discuss this in
more detail in~\Sec{subsec:VTP}. 

Before bosonizing the Coulomb interaction, we re-write the
$f$-electron occupation in terms of pseudospin-$\half$ operators,
$n^{f}_{j}-\half = \tau^{z}_{j}$. In the pseudospin representation,
spin-$\uparrow$ at site $j$ indicates an occupied $f$-orbital and
\emph{vice versa}. For the CP, the condition of constant  
$f$-electron concentration then translates into a fixed pseudospin
magnetization $m^{z} = n^{f}-\frac{1}{2}$. The use of the pseudospins
will considerably simplify the subsequent manipulations. We re-write
the Coulomb interaction  
\beq
G\sum_{j}n^{f}_{j}n^{c}_{j} =
G\sum_{j}\tau^{z}_{j}n^{c}_{j}
-\half{G}\sum_{j}\tau^{z}_{j} + \text{const.} \label{eq:B:Grewrite}
\eeq
We have used the requirement of constant total electron concentration to
obtain the second term. 
Substituting \eq{eq:B:b_nc} into \eq{eq:B:Grewrite}, we obtain the
bosonized form of the FKM Hamiltonian
\beqarray
\Ham[FKM] &= &\frac{\vF{a}}{2\pi}\sum_{j}\left\{\left(\partial_{x}\phi(x_j)\right)^{2}+\left(\partial_{x}\theta(x_j)\right)^{2}\right\}+G\left(n^{c}_{0}-\half\right)\sum_{j}\tau^{z}_{j} \notag \\
&&-\frac{Ga}{\pi}\sum_{j}\tau^{z}_{j}\partial_{x}\phi(x_j)+\frac{GAa}{\alpha}\sum_{\nu,j}\tau^{z}_{j}\Kfdag[\nu]\Kf[-\nu]e^{i2\nu\phi(x_j)}e^{-i2\nu{k}_{F}x_{j}}
\label{eq:B:HamFKM} 
\eeqarray
For $c$-electron concentration $n^{c}$, the Fermi velocity is defined
$\vF=-2ta\sin(\kF{a})$ where $\kF={\pi}n^{c}/a$.
Note that the parameter $\epsilon_{f}$ only enters
into~\eq{eq:B:HamFKM} indirectly through $\vF$ and $\kF$. 
Since the Klein factor products in the backscattering corrections
[the last term in~\eq{eq:B:HamFKM}] commute
with the Hamiltonian, we replace them by their expectation
value, $\Kfdag[\nu]\Kf[-\nu]=\langle\Kfdag[\nu]\Kf[-\nu]\rangle =1$.

\section{The Canonical Transform} \label{sec:CT}

The work on the CP has established that the $c$ electrons mediate
interactions between the $f$ electrons via the interorbital Coulomb
repulsion. Very little, however, is
known about the character of these interactions: here we seek
to reveal the electronic origins of the charge order by rotating the
Hilbert space basis to decouple the $c$ and  
$f$ electrons. We apply a lattice generalization of the canonical
transform used by Schotte and Schotte in the X-ray edge
problem (XEP):~\cite{SS69} 
\beq
\hat{U} =
\exp\left\{i\frac{Ga}{\pi\vF}\sum_{j'}\tau^{z}_{j}\theta(x_{j'})\right\} \label{eq:CT:CT}
\eeq
The canonical transform bears a close resemblance to the transform
used by Honner and Gul\'{a}csi in their analysis of the KLM.~\cite{HG}
This resemblance is not coincidental, but instead points to a
fundamental similarity between the FKM and KLM which we explore
below. 

A major advantage of the Bose representation is that the
transformation of the Bose operators under~\eq{eq:CT:CT} may be
calculated exactly using the Baker-Hausdorff
formula. This allows us to carry the canonical transform of the
Hamiltonian through to all orders. The effect of the transform may be
summarized as follows: 
\begin{eqnarray}
\hat{U}^{\dagger}\phi(x_j)\hat{U} &=& \phi(x_j)
-\frac{Ga}{2\vF}\sum_{j'}\tau^{z}_{j'}\sgn[\alpha]{x_{j'}-x_j} \label{eq:CT:CTphij}
\\
\hat{U}^{\dagger}\partial_{x}\phi(x_j)\hat{U} &=&
\partial_{x}\phi(x_j)+\frac{Ga}{\vF}\sum_{j'}\tau^{z}_{j'}\delta_{\alpha}(x_j-x_{j'}) \label{eq:CT:CTncj}
\end{eqnarray}
All other operators in \eq{eq:B:HamFKM} are unchanged by the
transform. In particular, we note that the transform preserves the
$f$-configuration,
i.e. $\hat{U}^{\dagger}\tau^{z}_{j}\hat{U}=\tau^{z}_{j}$. 

The transformation of the derivative of the $\phi$-field
[\eq{eq:CT:CTncj}] is of special note, as it makes 
explicit the dependence of the $c$-electron density  
$\rho(x_j) = n^{c}_{0}-\frac{a}{\pi}\partial_{x}\phi(x_j)$ at site $j$ 
upon the local $f$-electron occupation:
\beq
\hat{U}^{\dagger}\rho(x_j)\hat{U} = \rho(x_j) -\frac{Ga^2}{\pi\vF}\sum_{j'}\left(n^{f}_{j'}-\half\right)\delta_{\alpha}(x_j-x_{j'}) \label{eq:CT:rhoj}
\eeq
As expected, the effect of the Coulomb interaction is to enhance 
(deplete) the $c$-electron density where the $f$ orbitals are
empty (occupied). As we explain below, this is the origin of the
observed segregation in the CP.

Substituting the transformed Bose fields [\eq{eq:CT:CTphij} and
\eq{eq:CT:CTncj}] into \eq{eq:B:HamFKM}, we obtain
\beqarray
{\hat{U}}^{\dagger}\Ham[FKM]\hat{U}
&=&
\frac{\vF{a}}{2\pi}\sum_{j}\left\{\left(\partial_{x}\phi(x_j)\right)^{2}+\left(\partial_{x}\theta(x_j)\right)^{2}\right\} +G\left(n^{c}_{0}-\half\right)\sum_{j}\tau^{z}_{j}
\notag \\
&&
-\frac{G^2a^2}{2\pi\vF}\sum_{j,j'}\tau^{z}_{j}\delta_{\alpha}(x_j-x_{j'})\tau^{z}_{j'}
\notag \\
&& +\frac{2GAa}{\alpha}\sum_{j}\tau^{z}_{j}\cos\left(2\left[\phi(x_j)-{\cal{K}}(j)-\kF{x_j}\right]\right)
\label{eq:CT:HamFKM}
\eeqarray
where we have introduced the simplifying notation
\beq
{\cal{K}}(j) = \frac{Ga}{2\vF}\sum_{j'}\tau^{z}_{j'}\sgn[\alpha]{x_{j'}
- x_{j}} \label{eq:CT:calK}
\eeq
for the string operator in~\eq{eq:CT:CTphij}.
Since the canonical transformation of $\Ham[FKM]$ has been carried out
exactly, it follows that \eq{eq:CT:HamFKM} is identical to
\eq{eq:B:HamFKM}. The result of our transformation is to have
re-written $\Ham[FKM]$ in a new basis that includes the
effective interactions between the $f$ electrons. The rest of
this paper will be concerned with the study of \eq{eq:CT:HamFKM}; we
begin by examining the origins of the terms involving
the $f$ electrons in the transformed Hamiltonian.

\subsection{The Ising interaction} \label{subsec:SEGint}

The removal of the forward-scattering Coulomb interaction by the 
canonical transform introduces an effective
interaction between the $f$ electrons:
\beq
-\frac{G^2a^2}{2\pi\vF}\sum_{j,j'}\tau^{z}_{j}\delta_{\alpha}(x_j-x_{j'})\tau^{z}_{j'} \label{eq:CT:EffInt}
\eeq
Unlike other effective interactions, such as the weak-coupling RKKY
theory~\cite{SchlottmannSRO} or the large-$G$
expansion,~\cite{GJL92} \eq{eq:CT:EffInt} is
\emph{non-perturbative}. Furthermore, \eq{eq:CT:EffInt} differs from
these other effective interactions
in being responsible only for the segregation and phase separation. 
Its significance warrants some discussion upon its properties and
origins.

\begin{figure}[t]
\includegraphics[width=8cm]{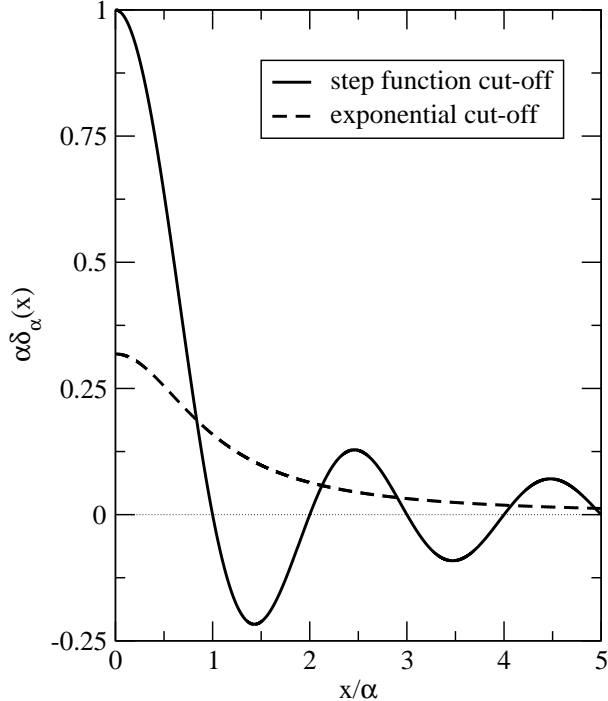} 
\caption{\label{fig:delta} The form of the interaction
  potential~\eq{eq:CT:EffInt} for step function 
  [$\CO{}=\Theta(|k|-\pi/\alpha)$] and exponential
  [$\CO{}=\exp\left(-\alpha|k|\right)$] cut-offs.}
\end{figure}

The interaction is implicitly dependent upon the properties of the
$c$ electrons: the form of the potential in~\eq{eq:CT:EffInt} is the
Fourier transform of the cut-off function
\beq
\delta_{\alpha}(x_{j}) =
\frac{1}{L}\sum_{k}\CO{}e^{ikx_{j}} \label{eq:CT:CO} 
\eeq
To concretely illustrate the variation of the interaction, we consider
two choices of cut-off
\beq
\CO{} = 
\begin{cases}
\Theta(|k|-\frac{\pi}{\alpha}) & \text{step function} \\
\exp\left(-\alpha|k|\right) & \text{exponential}
\end{cases}
\eeq
Here $\Theta(x)$ denotes the well-known Heaviside step function. 
For simplicity, we evaluate the summation~\eq{eq:CT:CO} in the
thermodynamic limit $L\rightarrow\infty$ and for a continuum system (valid for
$\alpha\gg{a}$). We thus 
find 
\beqarray
\delta_{\alpha}(x) & = & \int^{\infty}_{0}\frac{dk}{\pi}\cos(kx)\CO{}
\notag \\
&=& 
\begin{cases}
\sin(\pi{x}/\alpha)/(\pi{x}) &
\text{step function} \\
\pi^{-1}\alpha/(\alpha^2+x^2) & \text{exponential}
\end{cases} \label{eq:CT:smeareddelta}
\eeqarray
These integrals are plotted in~\fig{fig:delta}; this plot makes it clear that
$\alpha$ characterizes the range of the
interaction~\eq{eq:CT:EffInt}. For the step-function cut-off
the potential will take negative values for $x>\alpha$. Since
$\alpha$ is limited below by the lattice constant, however, the
nearest-neighbour value $\delta_{\alpha}(a)$ is always non-negative
and exceeds in magnitude all other values of the potential. In the pseudospin
language the interaction~\eq{eq:CT:EffInt} is 
ferromagnetic below the bosonic wavelength cut-off.  
Furthermore, beyond this length scale the potential is insignificant.

The canonical transform reveals that the forward-scattering
mediates attractive interactions between the $f$ electrons; as such,
it can account for the observed segregation~\cite{seg1D} and phase
separation.~\cite{FGM96,GJL96} This is not unexpected, as the
forward-scattering $c$ electrons transfer small crystal momentum
($\ll\kF$) to the $f$ orbitals, thus only interacting with the
long-wavelength features of the underlying $f$-electron configuration.  

To fully understand the physical origin of~\eq{eq:CT:EffInt} we
must consider the details of the bosonization process.
Because of the bosonic wavelength cut-off, our treatment
can only describe density fluctuations over distances
$>{\cal{O}}(\alpha)$. The bosonic fields cannot distinguish
separations less than this distance, hence the smeared canonical field
commutators~\eq{eq:B:comphitheta} and~\eq{eq:B:comdelphitheta}. Our
description of this system thus assumes that the $c$ electrons are
delocalized over a characteristic length $\sim\alpha$: 
the $\alpha$-smeared $\delta$-functions [\fig{fig:delta}] may be very crudely 
conceived as the probability density profile of these delocalized
electrons, i.e $|\psi(x)|^{2}\propto\delta_{\alpha}(x)$. 
The finite spread of the $c$-electron wavefunctions carries the
interorbital Coulomb repulsion over several lattice sites [see
\eq{eq:CT:rhoj}], directly leading to the segregating
interaction~\eq{eq:CT:EffInt}. 

In the familiar bosonization description of one-component systems such
as the Hubbard model, making $\alpha$ arbitrarily small does
not alter the critical properties of the model; in particular, we still obtain
the Luttinger liquid fixed point 
behaviour.~\cite{G:QP1D} In these models 
$\alpha$ is regarded as a short-distance cut-off which defines the
minimum length scale in the system (usually the average inter-electron
separation $\sim\kF^{-1}$), much as the infrared cut-off in field
theory. To understand 
the long-wavelength behaviour we need only keep $\alpha$ as a
formally finite parameter.
Such arguments cannot, however, be made for the $c$ electrons in the
FKM, where the limiting length scale of the bosonic description is
determined by the interactions with the localized $f$ electrons: the
parameter $\alpha$ therefore enters our bosonic theory as a finite
but undetermined length. We estimate $\alpha$ by
examining the short-range fermionic scattering of the $c$ electrons
off the $f$ orbitals.

As is well-known, the configuration adopted by the $f$ electrons in
the FKM acts as a single-particle potential for the $c$ electrons.
That is, the $c$ electrons move in a site-dependent potential 
that takes only two values $+G/2$ or $-G/2$, corresponding to occupied
and unoccupied underlying $f$ orbitals respectively. 
Below the average inter-electron separation ($\kF^{-1}=a/\pi{n^{c}}$), the
$c$ electrons move independently of one another and their 
motion is therefore described by a single-particle Schr\"{o}dinger
equation. In the limit $n^{c}\rightarrow0$ the
average inter-particle separation is much larger than 
the lattice constant: it is here acceptable to take the continuum
limit of the lattice model, yielding a simple form for the
Schr\"{o}dinger equation describing the low-energy ($E=0$)
wavefunctions $\psi(x)$:
\beq
\partial^{2}_{x}\psi(x) = Gm\langle{\tau^{z}_{x}}\rangle\psi(x)
\label{eq:CT:Schrodinger} 
\eeq
here $m$ is the bare electron mass. 

The motion of the $c$ electrons across the lattice is analogous to the
familiar problem of elementary quantum mechanics of a particle in a
finite well.~\cite{Schiff}  
For a $c$-electron moving in a region free from $f$ electrons
($\langle{\tau^{z}_{x}}\rangle=-\half$), the energy 
of the $c$ electron exceeds the potential and so we find the
solutions for $\psi(x)$ to be plane waves. In contrast, the
$c$-electron energy is less than the potential in a region of occupied
$f$ orbitals ($\langle{\tau^{z}_{x}}\rangle=\half$): we therefore find
exponentially decaying solutions $\psi(x)\sim\exp(-x/\zeta)$
characterized by the length scale $\zeta\sim\sqrt{1/G}$. Since we identify
$\alpha$ with the finite spread 
of the delocalized $c$ electrons across the orbitals, we conclude that
$\alpha\sim\zeta$. We therefore expect $\alpha = b\sqrt{t/G}$ where
$b$ is a constant to be determined.

\subsubsection{Relationship to the KLM} \label{subsubsec:KLM}

The similarity of the canonical
transform~\eq{eq:CT:CT} to that used by Honner and Gul\'{a}csi in
their treatment of the KLM suggests a connection between the two models.
This relationship is best revealed by considering the single-impurity limit 
of these lattice models; for the FKM, the associated impurity problem
is the XEP.

As is well known, the sudden appearance of the core hole in the XEP
excites an infinite number of electron-hole pairs in the conduction
band, leading to singular features in the X-ray spectrum (the orthogonality
catastrophe). Schotte and Schotte recast the problem in terms of
Tomonaga bosons: the core hole potential directly couples to the boson
modes of the scattering electrons, and may be removed by a suitable
shifting of the oscillator frequencies.~\cite{SS69} Our own canonical
transform~\eq{eq:CT:CT} repeats this procedure across the 1D
lattice. Although the $f$ electrons in the FKM are static, 
the appearance of the core hole in the XEP is equivalent to
suddenly turning on the interactions in the FKM. Since we start with 
non-interacting boson fields in~\eq{eq:B:HamFKM}, this is a perfect
analogy.  

The spin-$\half$ Kondo impurity is another classic example of the
orthogonality catastrophe, although the singular behaviour here arises
due to the shifting of the spin-sector boson frequencies. In
the usual Abelian bosonization approach the boson modes only
directly couple to the $z$-component of the impurity spin: ignoring
the transverse terms the problem is identical in form to the
XEP. Although these transverse terms somewhat complicate the
analysis, for special values of $J^{z}$ (the Toulouse
point),~\cite{T69} it is possible to map the problem to the exactly
solvable resonant-level model by shifting the $c$-electron boson
frequencies as in the XEP.~\cite{S70} For the lattice case this
argument may be generalized to arbitrary $J^{z}$:~\cite{HG} Honner and
Gul\'{a}csi's canonical transform therefore shifts the KLM's
spin-sector boson frequencies in precisely the same way as the
transform~\eq{eq:CT:CT} shifts the charge-sector boson frequencies in
the FKM.    

The similarity between the charge-sector physics of the FKM and the
spin-sector physics of the KLM suggests a parallel between the segregating
interaction~\eq{eq:CT:EffInt} and the Kondo double-exchange. This is made
explicit by our boson-pseudospin representation: ignoring the
backscattering term in~\eq{eq:B:HamFKM}, the Hamiltonian is identical 
to the spin-sector of a forward-scattering $J_{\perp}=0$ KLM. Within
their Abelian bosonization description, Honner and Gul\'{a}csi found
the forward-scattering $z$-exchange term in the KLM directly
responsible for mediating the double-exchange 
between the localized spins.~\cite{HG} The origin of this 
double-exchange term is therefore identical to our segregating
interaction.

The shifting of the FKM's charge-sector Bose frequencies produces 
distortions of the $c$-electron density in response to the local
$f$-occupation [see~\eq{eq:CT:rhoj}]. These deviations from the
homogeneous noninteracting density $n^{c}_{0}$ may be
interpreted as polaronic objects;~\cite{SS69} note however that
because of the lack of fluctuations in the $f$ orbitals, these
distortions are frozen into the ground state. This illustrates an
important departure from the KLM physics, where the spin-flip
($J_{\perp}$-) exchange terms cause the $z$-component of the lattice
spins to fluctuate, giving the distortions of the $c$-electron spin
density (i.e. spin polarons) mobility.

It is possible to modify the FKM in order to replicate this aspect of
the KLM physics. The simplest such extension is an on-site
hybridization term between the $c$ and $f$ orbitals,
$\Ham[hyb]=V\sum_{j}\{c^{\dagger}_{j}f^{}_{j}+\Hc\}$: adding 
$\Ham[hyb]$ to~\eq{eq:Intro:FKM} gives the quantum Falicov-Kimball model
(QFKM). Using a bosonization mapping at the Toulouse point,
Schlottmann found that the $J_{\perp}$-exchange term of the Kondo
impurity is equivalent to the hybridization potential in the
single-impurity limit of the QFKM.~\cite{Schlottmann} In the lattice
case, the polaronic distortions acquire mobility as in the KLM: this
coupling of the $c$- and $f$-electron densities may be identified as
a Toyozawa ``electronic polaron''.~\cite{T54} Electronic polarons
in the QFKM have previously been studied by Liu and Ho;~\cite{Liu&Ho}
our work on the 1D QFKM largely confirms their scenario.~\cite{BG06}
A complete account of this work is in preparation.~\cite{BGinprep}

\subsection{The longitudinal fields} \label{subsec:longitudinal}

The other two terms in the transformed Hamiltonian involving the
$\tau$ pseudospins are a constant and a site-dependent longitudinal
field. The former is only of importance to the VTP: the
renormalization of the $f$-level by the Coulomb interaction will drive
a ``classical'' valence transition. The sign of this term is proportional
to $n^{c}_{0}-\half$, which implies a strong dependence upon the
noninteracting band structure: if $\epsilon_{f}>0$, the $f$-level will
be renormalized upwards, emptying its contents into the $c$-band; for
$\epsilon_{f}<0$ the $f$-level is lowered below the
$c$-band, and so all electrons will eventually possess
$f$-electron character. Since this term does not determine the
configuration adopted by the $f$ electrons but rather only their
number, we leave further discussion to when we analyze the VTP
phase diagram.

The site-dependent field is of more general interest. This
originates from the $2\kF$-backscattering correction and is directly
responsible for the well-known crystalline order in the FKM. Before
demonstrating how the crystalline $f$-configurations can be extracted
from this term, we first briefly review the present understanding of
the weak-coupling periodic phases.

\begin{figure}[t]
\includegraphics[width=8cm]{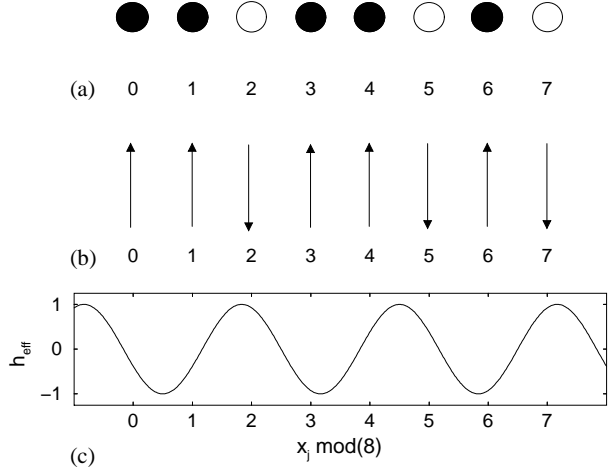} 
\caption{\label{fig:crystal} (a) The configuration of the $f$-ions in
  the weak-coupling homogeneous unit cell for $n^{c}=\frac{3}{8}$,
  $n^{f}=\frac{5}{8}$. The 
  filled and empty circles represent occupied and unoccupied sites
  respectively. (b) The pseudospin representation for the
  configuration (a). (c) The variation of the effective magnetic field 
  $h_{\text{\tiny{eff}}}$ produced by the $2\kF$-backscattering
  correction across the unit cell. $h_{\text{\tiny{eff}}}$ is in units
  of $2GAa/\alpha$.}
\end{figure}

The origin of the crystalline order is a
Peierls-like mechanism: a one-dimensional metal is always unstable
towards an insulating state when in the presence of a periodic
potential with wavevector $2\kF$.~\cite{P:QTS} 
In the context of the FKM, a Peierls instability can arise when the
$f$-electrons crystallize in a periodic configuration with wavevector
$2\kF$. This is the case for weak coupling and we repeat a
theorem due to Freericks and Falicov:~\cite{FF90} given rational
$c$-electron density $n^{c}=p/q$ ($p$ prime with respect to $q$) and
$G/t\ll{q}$, then for $f$-electron density $n^{f}=p^{f}/q$ ($p^{f}$ not
necessarily prime with respect to $q$) the $f$ electrons occupy the
sites $x = nq+k_{j}$ where $n$ is an arbitrary integer and the
$k_{j}$ satisfy the relation
\beq
(pk_{j})\bmod{q} = j, \quad j=0,1,\ldots,p^{f}-1 \label{eq:CT:crystalcond}
\eeq
For example, consider the case $n^{c}=\frac{3}{8}$ and
$n^{f}=\frac{5}{8}$. The unit cell has eight sites, and the
$f$ electrons occupy the first, second, fourth, fifth and seventh
sites [see~\fig{fig:crystal}(a)]. 

Our approach reproduces this important result. In the pure
crystalline phase, the $c$-electron spectrum is gapped.~\cite{F03} We 
therefore replace $\phi(x_{j})$ in the cosine term of~\eq{eq:CT:HamFKM} by the
uniform average $\langle{\phi}\rangle$. Ignoring the ${\cal{O}}(G^2)$
Ising interaction, at weak coupling the $\tau$ pseudospins are
therefore arranged by the field 
\beq
\frac{2GAa}{\alpha}\sum_{j}\tau^{z}_{j}\cos\left(2[\langle{\phi}\rangle+{\cal{K}}(j)-\kF{x_j}]\right) \label{eq:CT:LROfield}
\eeq
The string operator ${\cal{K}}(j)$ is a constant of the motion and so
it may be replaced by its eigenvalue. Referring to~\eq{eq:CT:calK}, 
this term subtracts the magnetization of the
$\tau$ pseudospins more than $\alpha$ to the right of site $j$ from
the magnetization of the $\tau$ pseudospins more than $\alpha$ to the
left of site $j$: for an infinite chain in the pure crystalline phase
this quantity vanishes, ${\cal{K}}(j)=0$. The value of
$\langle{\phi}\rangle$ 
must be chosen to minimize the backscattering energy. This of course
implies a non-trivial dependence upon the $f$-electron concentration:
for a $Q$-site unit cell with $f$-electron concentration $q^{f}/Q$, we
must minimize $\langle\phi\rangle$ over the sum 
\beq
\sum_{l=1,q^{f}}\lambda_{l} - \sum_{l=q^{f}+1,Q}\lambda_{l}
\label{eq:CT:minsum} 
\eeq
where $\lambda_{l}=\cos\{2[\langle{\phi}\rangle-\kF{x_{l}}]\}$ and
$l$ is chosen such that
$\lambda_{1}\leq\lambda_{2}\leq\ldots\leq\lambda_{Q}$, $x_{l}$ lying
within the unit cell. 
This minimization is most easily accomplished numerically;
$\langle\phi\rangle$ is restricted to values in the
interval $[0,\pi)$. 

The sum~\eq{eq:CT:minsum} assumes that the $q^{f}$ $f$ electrons per
unit cell will occupy the $q^{f}$ lowest-energy sites in the
potential~\eq{eq:CT:LROfield}. In terms of the pseudospins, there is a
fixed magnetization $q^{f}/Q-\frac{1}{2}$ per unit cell; the
$\uparrow$-spins occupy the sites with the smallest magnetic field,
with the $\downarrow$-spins 
sitting on the remaining $Q-q^{f}$ sites. 
For the example above with $n^{c}=\frac{3}{8}$ and
$n^{f}=\frac{5}{8}$, we find $\langle\phi\rangle\approx0.589$ and so
the $f$ electrons experience a potential  
$$
\frac{2GAa}{\alpha}\sum_{j}\tau^{z}_{j}\cos\left(1.178-\frac{3\pi{x_{j}}}{4a}\right)
$$
We plot this potential in~\fig{fig:crystal}(c) along with the
$\tau$ pseudospin orientations [\fig{fig:crystal}(b)]. It is in good
agreement with the exact 
result~\eq{eq:CT:crystalcond}, although there is some ambiguity with
respect the position of the $f$-electron at the fifth and sixth sites.
Closer correspondence may
be achieved by taking into account higher-order backscattering
processes; because bosonization is fundamentally a long-wavelength 
method, this approach does not replace the exact
calculations. Nevertheless, our analysis convincingly demonstrates
that bosonization is capable of describing crystallization of the
$f$ electrons in the FKM. 

Before proceeding to a discussion of the FKM's phase diagram, we note that as
the 
number of $f$ electrons is limited in the FKM, it may not be
possible for a pure crystalline phase to gap the $c$-electron spectrum
at the Fermi energy. In particular, for irrational $c$-electron
filling the field~\eq{eq:CT:LROfield} will be incommensurate with the
lattice. Although this situation remains unclear, in the related case
of rational $c$-electron filling $n^{c}=p/q$ and $f$-electron
filling $p'/q<n^{f}<(p'+1)/q$ the system phase separates into 
regions with periodic phases determined by~\eq{eq:CT:crystalcond} for
$p^{f}=p'$ and $p^{f}=p'+1$.~\cite{FGM96} It is also known that for 
$f$-electron concentrations $n^{f}\lesssim0.371$ and
$n^{f}\gtrsim0.629$ the system exists in a mixture of a crystalline
phase and a homogeneous phase (the empty or full configuration).
This phase separation behaviour cannot be explained purely in terms of
$c$-electron backscattering.   

\section{The Effective Hamiltonian} \label{sec:EH}

By themselves, the Ising interaction~[\eq{eq:CT:EffInt}] and the
longitudinal field~[\eq{eq:CT:LROfield}] explain the SEG
and crystalline phases respectively. To understand the origin of the
phase separation or interpret the numerically-determined phase
diagram, however, we must consider the interplay of these terms. In
particular, it is desirable to have a simple effective
Hamiltonian for the $f$ electrons that includes both the
crystallizing and segregating tendencies of the FKM.

The transformed Hamiltonian~\eq{eq:CT:HamFKM} offers a
straight-forward route to such a description of the $f$-electrons. 
With the removal of the term describing the forward-scattering interaction,
the only coupling between the two species is in the $2\kF$-backscattering
correction. In the 
weak-coupling crystalline phases it is
possible to completely decouple the $f$ electrons from the
$c$ electrons by replacing the bosonic $\phi(x_j)$ field by its
expectation value. Combining~\eq{eq:CT:LROfield}
with the interaction~\eq{eq:CT:EffInt}, we obtain an effective
spin-$\half$ Ising model for the $f$ electrons valid throughout the
region where the crystalline phases are realized:
\beq
\Ham[eff] =
-\frac{G^2a^2}{2\pi\vF}\sum_{j,j'}\tau^{z}_{j}\delta_{\alpha}(x_{j}-x_{j'})\tau^{z}_{j'}
+
\frac{2GAa}{\alpha}\sum_{j}\tau^{z}_{j}\cos\left(2[\langle{\phi}\rangle+{\cal{K}}(j)-\kF{x_j}]\right) \label{eq:EH:EffHam}
\eeq

This is an important result, but our approach is not limited only to
the crystalline phases: for other configurations, the form of
the effective Hamiltonian~\eq{eq:EH:EffHam} remains valid, although the 
site-dependence of the longitudinal field is different. In the
crystalline phases ${\cal{K}}(j)$ is vanishing; in the
segregated or phase separated states, however, ${\cal{K}}(j)$ has linear
variation. Ignoring the short-range deviation of $\sgn[\alpha]{x_j}$ from the
true sign-function, we write 
\beq
{\cal{K}}(j) \approx
\frac{Ga}{\vF}\sum_{n=1}\left(\tau^{z}_{j+n}-\tau^{z}_{j-n}\right) 
\label{eq:EH:Kj} 
\eeq
Assume a phase separation between phase  $A$ and phase $B$ with the
boundary at $j=0$. Then for $j'\gg{1}$ we have approximately~\cite{G04}
\beq
{\cal{K}}(j') \approx {\cal{K}}(0) +
\frac{Ga}{\vF}\left(\langle\tau^{z}\rangle_{A} -
\langle\tau^{z}\rangle_{B}\right)|j'| \label{eq:EH:approxKj}
\eeq
where the subscripts $A$ and $B$ refer to the magnetization in the
$A$ and $B$ phases respectively. We have chosen the sign
of the linear term by choosing phase $A$ to be realized to the
right and phase $B$ to the left of $j=0$. Note that
${\cal{K}}(j)$ is constant for a pure phase, as we expect.

For the SEG phase, the division of the lattice into empty and full sections
implies a variation ${\cal{K}}(j)\sim{(Ga/2\vF)|j|}$. Although the conduction
electron spectrum does not display a gap, the decoupling procedure for the
field $\phi(x_j)$ used in~\Sec{subsec:longitudinal} may be easily
generalized. In the SEG phase, the conduction electrons are restricted
to a fraction $(1-n^{f})$ of the lattice, where they behave as a
non-interacting electron gas. We therefore replace $\phi$
in~\eq{eq:CT:LROfield} by its non-interacting average
$\langle{\phi}\rangle=0$ to obtain the effective pseudospin Hamiltonian in the
SEG phase.

The phase-separation between the crystalline and empty phase is the
most complex situation to analyze, as we must account for the very
different behaviour of the $c$-electrons for the two configurations.
Exact diagonalization calculations on 3200 site chains reveal that the
momentum distribution of the 
$c$-electrons is essentially a superposition of the gapped and
noninteracting forms corresponding to the crystalline and empty
regions of the lattice, with vanishingly small correction due to the
interface of these phases in the thermodynamic limit
$N\rightarrow\infty$.~\cite{F03}
Decoupling the $c$-electron fields as above, we 
take different averages of the $\phi$-field in the bulk of the two
phases: for the empty phase, we use the noninteracting value
$\langle\phi\rangle=0$ while we determine $\langle\phi\rangle$ for the
crystalline phase as in~\Sec{subsec:longitudinal}. Approximating
${\cal{K}}(j)$ as in~\eq{eq:EH:approxKj}, we have
${\cal{K}}(j)\sim{(Ga/2\vF)m|j|}$ where 
$m=\langle\tau^{z}\rangle_{P}+\half$ ($\langle\tau^{z}\rangle_{P}$
is the magnetization of the periodic phase). 

The effective Hamiltonian across the phase diagram is 
ferromagnetic Ising model in a oscillatory longitudinal
field. This model exhibits all the
most important aspects of FKM physics. It is, however, 
important to add here a cautionary note about the limitations
of~\eq{eq:EH:EffHam}.   
Bosonization is an inherently long-wavelength method, and so it is
therefore unreasonable to expect $\Ham[eff]$ to precisely reproduce
the microscopic details of the $f$-electron configuration realized for
given $G$, $n^c$ and $n^f$. Rather, $\Ham[eff]$ is primarily relevant to the
long-wavelength physics, with the site-dependent longitudinal field acting as
an essentially approximate account of the short-range crystallizing
interactions. 
Furthermore, the form of $\Ham[eff]$ is rigorously quantitatively
valid only for $G\lesssim{t}$. Nevertheless, we expect that $\Ham[eff]$
is at least qualitatively correct over a much larger region of the phase
diagram.~\cite{G04} 

\section{The ground state phase diagram} \label{sec:PD}

\subsection{The crystallization problem} \label{subsec:CP}

In the CP, the concentration of the $f$-electrons is fixed: the
problem of the ground state phase diagram is then reduced to finding
the pseudospin configuration with magnetization $n^{f}-\half$ that
minimizes the energy $\langle{\Ham[eff]}\rangle$. This lattice-gas
problem, although conceptually simple, does not have a general solution. It is
therefore appropriate to use approximate methods to understand the physics.

In general, the segregating Ising interaction has a range $\alpha$
that extends over several lattice sites. To understand the segregation,
however, we need consider only the nearest-neighbour value of the potential 
$\delta_{\alpha}(a)$. That is, we write the Ising interaction
\beq
-\frac{G^2a^2}{2\pi\vF}\sum_{j,j'}\tau^{z}_{j}\delta_{\alpha}(x_{j}-x_{j'})\tau^{z}_{j'}
\approx
-\frac{G^2a^2}{\pi\vF}\delta_{\alpha}(a)\sum_{j}\tau^{z}_{j}\tau^{z}_{j+1} \label{eq:PD:nnapprox}
\eeq
This is justified as for realistic cut-off $\CO{}$ the interaction
potential $\delta_{\alpha}(x)$ is attractive for $x<{\alpha}$ but
falls off very quickly with distance [see~\fig{fig:delta}]. Truncating the
interaction should not significantly alter the critical properties of
the model, while considerably simplifying the analysis.

The weak-coupling phase separation into the empty and a periodic
configuration requires us to extend the Ising interaction beyond the
nearest-neighbour approximation used above. The restriction to fixed
magnetization makes this a challenging problem to analyze and a general
criteria for the phase separation is beyond the capabilities of our 
approach. We shall nevertheless demonstrate the origin of the phase
separation for a single set of input parameters, noting the importance
of considering a delocalization length $\alpha>2a$. 

\subsubsection{Segregation} \label{subsubsec:seg}

In the pseudospin ``language'' of the effective Ising
model~\eq{eq:EH:EffHam}, the SEG phase corresponds to a
ferromagnetic state with two domains: a single block of $\uparrow$-spins
occupying a fraction $n^{f}$ of the lattice and $\downarrow$-spins in
the remaining $(1-n^{f})N$ sites. From the form of $\Ham[eff]$, we see
that the critical Coulomb repulsion $G_{c}$ for the onset of
segregation is related to the
critical ratio $J/h$ for the onset of ferromagnetism in the model
\beq
\Ham = -J\sum_{j}\tau^{z}_{j}\tau^{z}_{j+1} +
h\sum_{j}\tau^{z}_{j}\cos(\omega_{j}{j}+\phi_{j}) \label{eq:PD:Ising}
\eeq
where $\omega_{j}$ and $\phi_{j}$ take different constant values in
different macroscopic regions of the lattice. 

The Ising model~\eq{eq:PD:Ising} has been studied for constant
$\omega$ and $\phi$ by Sire~\cite{S93}. For $\omega/\pi$ irrational,
i.e. the quasiperiodic Ising Model (QPIM), it is found that the critical
Ising coupling has the form
$J_{c}=h/\sin(\half{\omega})$. At couplings $J>J_{c}$ the ground 
state is ferromagnetically ordered; the adiabatic phase (where the
spins align antiparallel to the direction of the longitudinal field
$h^{z}_{j}$) is however only realized for $J<J_{c2}<J_{c}$ where
$J_{c2}=h\sin(\half\eta\omega)\sin(\half[\eta+1]\omega)/\sin(\half\omega)$
and $\eta$ is the largest integer smaller than $\pi/\omega$.
For the intermediate
couplings $J_{c}>J>J_{c2}$ the ground state is a ``mixture'' of the
adiabatic and ferromagnetic phases. This ``mixed'' state consists of a
quasiperiodic arrangement of clusters of adiabatically- and 
ferromagnetically-ordered spins. These clusters form as neither the
Ising interaction nor the magnetic field are strong enough to order the
entire lattice: ferromagnetic clusters occur where the magnetic
field is weak compared to the Ising term, while paramagnetic clusters are found
where the Ising term is weak compared to the magnetic field. Note
that this is {\it{not}} a phase separation phenomenon.  

The work on the QPIM in~\Ref{S93} was performed within the grand canonical
ensemble and 
so we must be cautious in relating these results to the effective
Hamiltonian~\eq{eq:EH:EffHam}. The expression for the critical
Ising coupling $J_{c}$ was deduced from
general arguments that should remain valid at fixed magnetization.
Indeed, the difference in energy per site between the
single-domain and the two-domain (SEG) solution vanishes as ${\cal{O}}(N^{-1})$
in the thermodynamic limit. The QPIM results should therefore correctly 
capture the competition between the adiabatic and ferromagnetic
orders present in~\eq{eq:EH:EffHam}: this provides a 
condition for segregation to dominate crystallization. 
Although the QPIM at weak- and strong-coupling corresponds 
to the behaviour seen in the small- and large-$G$ FKM, the agreement
breaks down at intermediate coupling. This is due to
the use of the grand canonical ensemble in~\Ref{S93}, as phase
separation cannot occur in the FKM without fixed electron
concentration~\cite{GUJ94}.  

We note that the effective
Ising Hamiltonian derived for the SEG phase must always display
ferromagnetic order. That is, within its range of applicability, the
Ising interaction always dominates the magnetic field. Although the
SEG effective Hamiltonian might display adiabatic order at weak
coupling, since crystallization in the FKM occurs in this limit a
different form of the longitudinal field must be used
in~\eq{eq:EH:EffHam}. To use the QPIM condition to determine the boundary
of the SEG phase, we therefore assume that the range of applicability of
the SEG effective Hamiltonian corresponds exactly to the extent of
ferromagnetic order.

A further difficulty encountered when applying the FM condition derived for
the QPIM 
is that the frequency of the magnetic field in the SEG phase takes two
values $\omega_{\pm}\approx{2(\pi{n^{c}}\pm{Ga/2\vF})}$ for each bulk
phase (i.e. the empty and full sections of the lattice). Although 
it is not possible to determine which value is realized for which section, 
the FM condition also holds for half-spaces and so we choose
$\omega_{+}$ which gives the observed monotonic dependence of the
critical line on the filling for weak- and intermediate-coupling~\cite{GJL96}. 
Since bosonization is quantitatively correct in the weak-coupling
limit, the expressions for $\omega_{+}$ will only rigorously hold 
for $G$ small as compared to the conduction electron bandwidth. In this limit
segregation occurs for $n^{c}\rightarrow0$ and so $\omega_{+}\approx{Ga/\vF}$;
we use this form to determine the weak-coupling critical line. 

Comparing the coefficients in~\eq{eq:EH:EffHam} with those
in~\eq{eq:PD:Ising} we find after some algebra the
condition for segregation 
\beq
\lim_{n^{c}\rightarrow0}\frac{G_{c}a}{\vF}\sin(\omega_{+}/2)=
\frac{G_{c}a}{\vF}\sin({G_{c}a}/{2\vF}) =
\frac{2A\pi}{\alpha\delta_{\alpha}(a)} 
\label{eq:PD:SEGcond} 
\eeq
We immediately deduce an important feature of the phase diagram: 
from general principles  
we know that $\alpha\gtrsim{\cal{O}}(\kF^{-1})$ as
$n^{c}\rightarrow0$. Since $\delta_{\alpha}(a)\sim{\alpha^{-1}}$ for
$\alpha\gg{a}$, the denominator of the RHS
of~\eq{eq:PD:SEGcond} tends to a constant as
$\alpha\rightarrow\infty$. For the expression to be consistent, we
hence require $G_{c}a/\vF=\mbox{constant}>0$ as
$n^{c}\rightarrow0$: we recover the result that segregation occurs at
arbitrarily small $G$ in the limit of vanishing conduction electron
concentration~\cite{KL86}.

\begin{figure}[t]
\includegraphics[width=8cm]{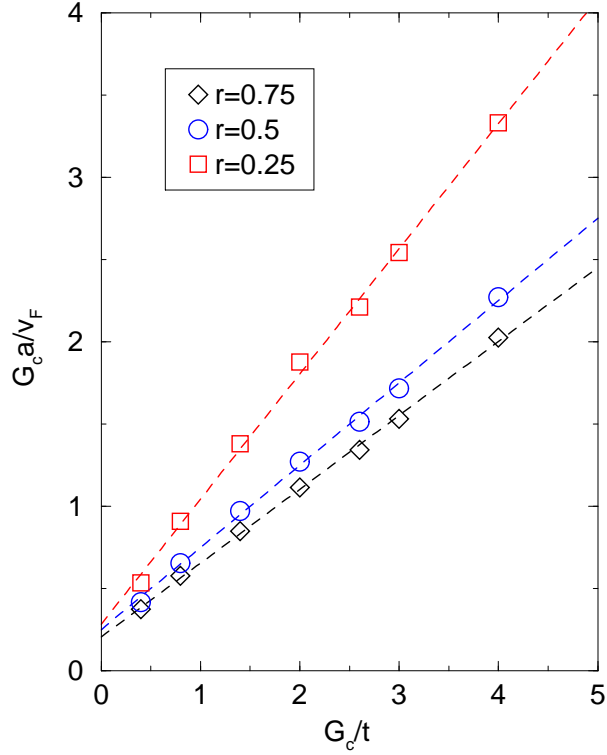} 
\caption{\label{fig:CP1} (color online) Dependence of $G_{c}a/\vF$ on
  $G_{c}/t$ for three values of the ratio $r$. The data is taken
  from~\Ref{GJL96}.}  
\end{figure} 

For finite filling, we can use our estimate for
$\alpha/a\sim{\sqrt{t/G}}$ to obtain the critical
line $G_{c}$ at weak- to intermediate-coupling. Assuming
$G_{c}a/\vF\ll{\pi/2}$ for small $G_{c}$, we linearize the sine function
in~\eq{eq:PD:SEGcond}; after some algebra we find
\beq
G_{c} =
4t\sin(\pi{n^{c}})\sqrt{\frac{A\pi}{\alpha\delta_{\alpha}(a)}}
\label{eq:PD:smalln} 
\eeq
At weak coupling, we have $\alpha\gg{a}$: Taylor-expanding the RHS
of~\eq{eq:PD:smalln} in powers of $a/\alpha$, we keep terms up to
second order. The coefficients in this expansion are dependent upon
the form of $\CO{}$ used; for exponential cut-off we have
\beq
G_{c}a/\vF \approx \sqrt{4A\pi^2}\left(1+\frac{a^2}{2\alpha^2}\right)
\label{eq:PD:GcTaylor} 
\eeq
Substituting our estimate for $\alpha$ into this equation, we thus
expect a linear relationship between $G_{c}a/\vF$ and $G_{c}$. This
also holds at intermediate couplings, as
clearly verified by the numerical results of~\Ref{GJL96}: we plot
$G_{c}a/\vF$ as a function of $G_{c}$ in~\fig{fig:CP1} for three
values of the fraction of electrons in the $c$-band $r=n^{c}/n=0.75$,
$0.5$ and $0.25$. 
After some re-arrangement of~\eq{eq:PD:GcTaylor}, we obtain the
general form of the critical line
\beq
G_{c}(r,n) = \frac{2B(r)\sin(\pi{rn})}{1-2C(r)\sin(\pi{rn})} \label{eq:PD:Gc}
\eeq
The numerical constants $B(r)$ and $C(r)$ are the $y$-intercept and gradient
of the lines in~\fig{fig:CP1} respectively; they are related to the
fitting parameters in~\eq{eq:PD:SEGcond} by $A={B(r)}^2/4\pi^2$ and
$\alpha/a=\sqrt{B(r)t/2C(r)G_c}$.

\begin{figure}[t]
\includegraphics[width=8cm]{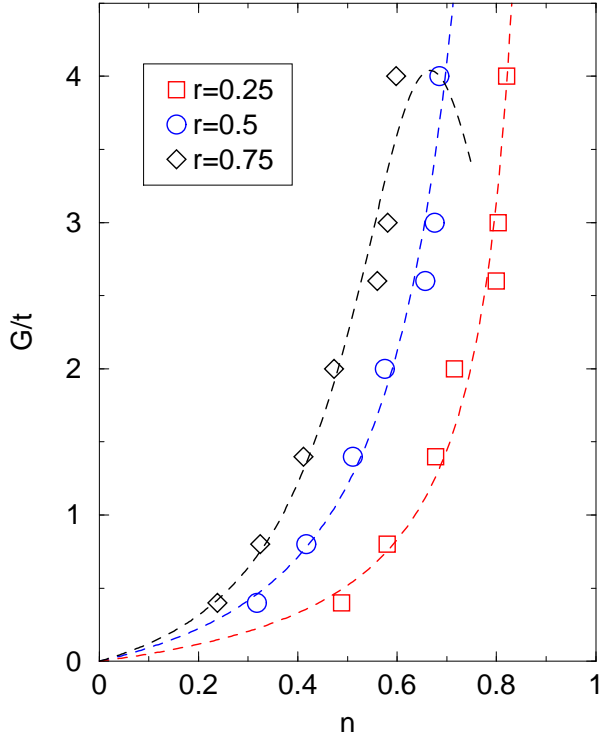} 
\caption{\label{fig:CP2} (color online) Dependence of $G_{c}/t$ on
$n$ for three values of the ratio $r$. The data is taken
from~\Ref{GJL96}. The critical lines $G_{c}(n)$ of best fit 
are as derived from~\fig{fig:CP1}. For each value of $r$, the SEG
phase occurs for $G>G_{c}(n)$.}
\end{figure}

From our fit to the lines in~\fig{fig:CP1} we find the critical lines
for the three values of $r$ 
\beq
G_{c}(n)/t = 
\begin{cases}
\vspace{0.25cm}
\displaystyle{\frac{0.5666\sin(n\pi/4)}{1-1.5212\sin(n\pi/4)}} & r=0.25 \\
\vspace{0.25cm}
\hspace{0.5cm}\displaystyle{\frac{0.5\sin(n\pi/2)}{1-\sin(n\pi/2)}} & r=0.5 \\
\displaystyle{\frac{0.4112\sin(3n\pi/4)}{1-0.8982\sin(3n\pi/4)}} & r=0.75
\end{cases} \label{eq:PD:CPGc}
\eeq
These three curves, along with the numerical data, are plotted
in~\fig{fig:CP2}. The curves track the data very well 
for both $r=0.25$ and $r=0.5$; for $r=0.75$, however, there is a
significant divergence between~\eq{eq:PD:CPGc} and the numerical
results as the coupling increases. The curve~\eq{eq:PD:CPGc} has a maximum
at $n=2/3$ (i.e. $n^{c}_{0}=1/2$), but no evidence of this maximum is found in 
the weak-coupling numerical results. Rather, we expect the critical
line to continue to diverge as half-filling is approached. We thus
conclude that there is a change in the form of $G_{c}(n)$ at
intermediate-coupling.
The numerical analysis of Gruber \emph{et al.} indicates that this
occurs at approximately $G\approx2.5t$~\cite{GUJ94}, which is
consistent with the observed deviation from the weak-coupling critical
line in~\fig{fig:CP2}.  

Any deviation from the weak-coupling form~\eq{eq:PD:Gc} for $r=0.25$
and $r=0.5$ is much less obvious. Since the $G\rightarrow\infty$
asymptotic form of $G_{c}(r,n)$ stated in~\Ref{GUJ94} is not
the same as that given by~\eq{eq:PD:CPGc}, we do however expect that a
different expression is valid at $G\gg{t}$. A new
functional dependence on $n$ in the strong-coupling regime is 
reasonable and does not contradict our own analysis: we have emphasized that
bosonization is only quantitatively accurate for weak-coupling. Importantly,
the physical processes driving the segregation will remain invariant across
the phase diagram. 

\subsubsection{Phase separation} \label{subsubsec:PS}

At weak-coupling and sufficiently small or large $f$-electron
concentration, the FKM is unstable towards a phase
separation between a homogeneous and a crystalline
configuration.~\cite{FGM96} Unlike the SEG phase, it is necessary to consider
the range of the forward-scattering Ising interaction as extending beyond the 
nearest neighbour to observe these phases.

To illustrate the importance of these long-range terms, we
examine the weak-coupling limit of~\eq{eq:EH:EffHam} for the case
$n^{c}=\half$, $n^{f}=\frac{1}{4}$. Of the Ising interaction~\eq{eq:CT:EffInt}
we keep the nearest-neighbour ${\cal{J}}_{1}$ and next-nearest
neighbour ${\cal{J}}_{2}$ terms. For $G\ll{t}$, we discard
${\cal{K}}(j)$ in the cosine's argument, leaving a 
staggered-field variation. We thus find an effective Hamiltonian of
the form 
\beq
\Ham = -{\cal{J}}_{1}\sum_{j}\tau^{z}_{j}\tau^{z}_{j+1} -
     {\cal{J}}_{2}\sum_{j}\tau^{z}_{j}\tau^{z}_{j+2} -
     h\sum_{j}(-1)^{j}\tau^{z}_{j} \label{eq:PD:nnnIsing}
\eeq
We calculate $E=\langle{\Ham}\rangle$ for three situations: (a) the
most homogeneous phase with period-$4$ pseudospin unit cell 
$[\uparrow\downarrow\downarrow\downarrow]$; (b) phase separation
between the empty phase ($[\downarrow]$) and the period-$2$ phase with
unit cell $[\uparrow\downarrow]$; and (c) segregation. We find the energy per
site for each of these configurations
\beq
E/N = 
\begin{cases}
-\tfrac{1}{4}{h} & \text{config. (a)} \\
-\half{\cal{J}}_{2}-\tfrac{1}{4}{h} & \text{config. (b)} \\
-\half{\cal{J}}_{1}-\half{\cal{J}}_{2} & \mbox{config. (c)}
\end{cases} \label{eq:PD:energies}
\eeq
These expressions hold in the thermodynamic limit
$N\rightarrow\infty$. 

We see that even an arbitrarily small
${\cal{J}}_{2}$ destabilizes configuration (a)
toward phase separation. Since at weak-coupling we expect
${\cal{J}}_{1}\ll{h}$, segregation will however not occur; instead 
configuration (b) is the most stable. This is the weak-coupling
phase separation between a crystalline and the empty phase found by
Freericks \emph{et al}.~\cite{FGM96} Although our analysis does not
provide a general condition for this peculiar form of phase separation
to occur, it shows that the physical origin of this effect is 
the competition between segregation and crystallization. We also see
how fixed $c$- and $f$-electron populations are essential for the
appearance of this phenomenon.

The phase separation between periodic and uniform states in
the $n<1$ FKM is not confined to the weak-coupling limit of the phase
diagram, but is also present at intermediate- and
strong-coupling.~\cite{FB93,GJL96} For $G>t$ our bosonization approach will at
least qualitatively capture the physics of the FKM: we therefore expect that
the competition between segregation and crystallization that we have
identified as the origin of the weak-coupling phase separation will
also be responsible for these intermediate- and strong-coupling phases.

\subsection{The valence transition problem} \label{subsec:VTP}

In contrast to the CP, the VTP has received little attention in the
FKM literature, despite the two interpretations being very closely related. In
both the CP and the VTP the $f$-orbital occupation is a good quantum number,
and the ground state may be defined as the configuration of the $f$ electrons
that minimizes the energy of the $c$ electrons. 
The only difference between the CP and the VTP is that in
the former the distribution of the electrons across the orbitals is
fixed, while in the latter the interactions determine the equilibrium
populations. 

For given interacting equilibrium populations in the VTP, the
configuration adopted by the $f$-electrons should be the same as in
the CP for the same fixed electron populations. 
As discussed in~\Sec{subsec:longitudinal}, the first two terms
of~\eq{eq:CT:HamFKM} determine the equilibrium 
distribution of the electrons across the $c$ and $f$
orbitals. They can be identified as the noninteracting $c$-electron
Hamiltonian and the $f$-level shift due to the Coulomb
repulsion. To estimate the
electron distribution for finite $G$ we therefore assume that the distribution
of the $nN$ electrons across the two orbitals in the FKM is the same as
in the system
\beq
{\cal{H}} = -t\sum_{j}\left\{c^{\dagger}_{j}c^{}_{j+1}+\Hc\right\} +
[\epsilon_{f}+G(n^{c}_{0}-\half)]\sum_{j}n^{f}_{j}
\eeq
That is, the contribution of the ordering terms in~\eq{eq:CT:HamFKM}
to the shift in electron density between the $c$ and $f$ orbitals is
taken to be negligible. 
This can be easily justified for a thermodynamically large system: the
difference between the energy per site of the SEG 
phase and the empty or full phases due to the Ising interaction is of order
$1/N$; and the average value of the bacscattering longitudinal field across
the lattice is vanishing.

We find that for noninteracting
$c$-electron population $n^{c}_0$ (fixed by the band structure)
the $c$-electron population in the interacting system is given by 
\beq
n^{c}=\frac{1}{\pi}\arccos\left(\cos(n^{c}_{0}\pi)-\frac{G}{4t}(2n^{c}_{0}-1)\right) \label{eq:PD:VTcond1}
\eeq
In~\Ref{GJL96} phase diagrams for the CP in the $n^{c}$-$n^{f}$ plane at
constant $G$ are presented For each $G$ the 
boundary between the SEG phase and the crystalline or phase separated
states is given by a straight line of the form
${n^{c}}=\gamma(1-n^{f})$ where $\gamma$ is a constant determined from
the numerical phase diagrams. Using the fixed electron concentration
condition we may re-write this  
\beq
n^{c}=\frac{\gamma}{1-\gamma}(1-n) \label{eq:PD:VTcond2}
\eeq
where $n=n^{c}+n^{f}$ is the total electron concentration. Equating
the RHS of~\eq{eq:PD:VTcond1} and~\eq{eq:PD:VTcond2} we obtain the
equation
\beq
\cos\left(r_{0}n\pi\right)-\frac{G}{4t}\left(2r_{0}n-1\right) -
\cos\left(\frac{\pi\gamma}{1-\gamma}[1-n]\right)=0 \label{eq:PD:VTcond3}
\eeq
where $r_0=n^{c}_{0}/n$ is the fraction of $c$-electrons 
in the non-interacting system. For given $r_0$ and $G$,
the value of $n$ that solves~\eq{eq:PD:VTcond3} gives the maximum
filling for which the SEG phase is stable. 

\begin{figure}[t]
\includegraphics[width=8cm]{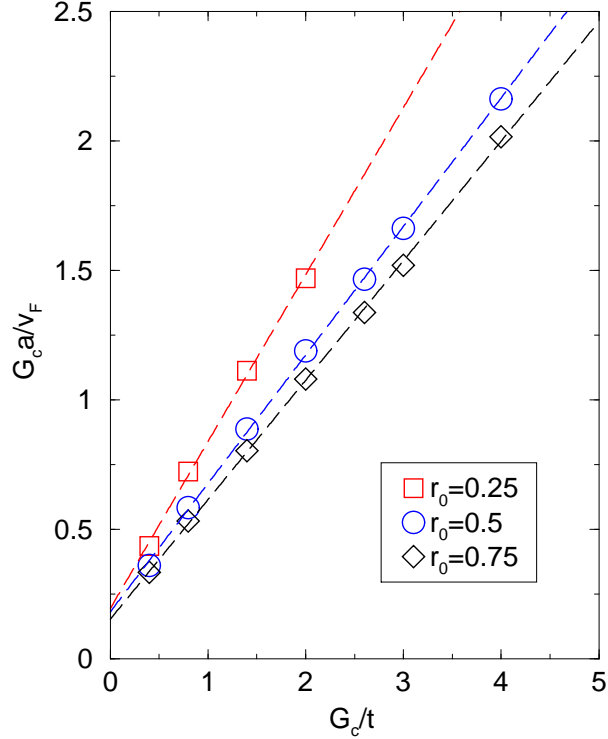} 
\caption{\label{fig:VTP1} (color online) Dependence of $G_{c}a/\vF$ 
  on $G_{c}/t$ for three values of the ratio $r_{0}$ in the VTP. The data is
  taken from~\Ref{GJL96}.}
\end{figure} 

\begin{figure}[t]
\includegraphics[width=8cm]{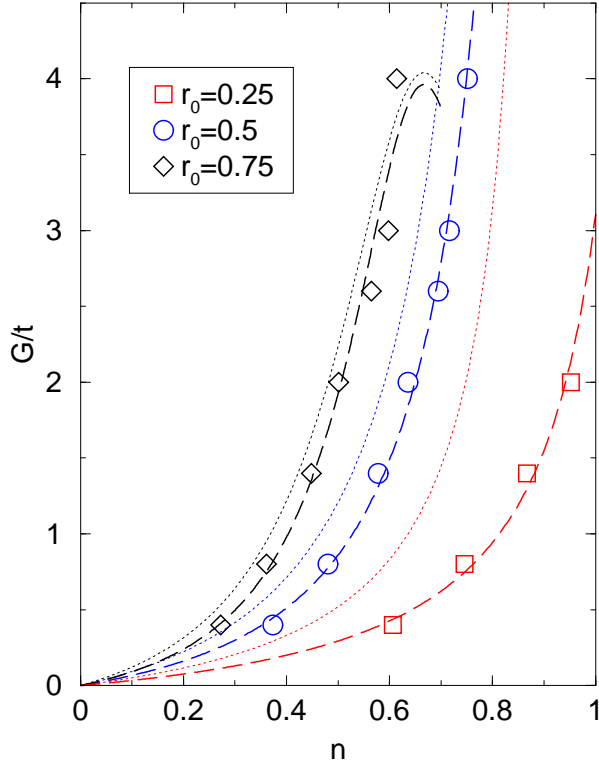} 
\caption{\label{fig:VTP2} (color online) Dependence of $G_{c}/t$ 
on $n$ for three values of the ratio $r$ for the VTP. The data is
taken from~\Ref{GJL96}. The critical lines
$G_{c}(n)$ of best fit (thick dashed lines) are as derived
from~\fig{fig:VTP1}. For each value of $r_{0}$, the SEG phase occurs for
$G>G_{c}(n)$. The thin dotted lines are the critical lines in the
$r=r_{0}$ CP.} 
\end{figure}

Using this procedure we calculate from the results of~\Ref{GJL96} the critical
value of $n$ for $r_{0}=0.25$, $0.5$ and  
$0.75$. Proceeding with our analysis as in the CP
(\Sec{subsubsec:seg}) we find that the linear relationship between
$G_{c}a/\vF$ and $G_{c}$ is also well obeyed in the VTP
[\fig{fig:VTP1}]. As such, at weak-coupling the critical line
$G_{c}(r_{0},n)$ has the form given by~\eq{eq:PD:Gc}. In
particular, we find from the linear best fit to the data
in~\fig{fig:VTP1} the following expressions
\beq
G_{c}(n)/t = 
\begin{cases}
\vspace{2.5mm}
\displaystyle{\frac{0.389\sin(n\pi/4)}{1-1.289\sin(n\pi/4)}} & r_0=0.25 \\
\vspace{2.5mm}
\displaystyle{\frac{0.363\sin(n\pi/2)}{1-0.992\sin(n\pi/2)}} & r_0=0.5 \\
\displaystyle{\frac{0.307\sin(3n\pi/4)}{1-0.923\sin(3n\pi/4)}} & r_0=0.75
\end{cases} \label{eq:PD:VTPGc}
\eeq
These are illustrated in~\fig{fig:VTP2} along with the critical lines
for the associated $r=r_0$ CP [\eq{eq:PD:CPGc}]. As before, we find
excellent agreement between the data points and the fitted curves for
both $r_0=0.5$ and $r_0=0.25$. Again, however, we find for
$r_{0}=0.25$ a significant divergence between the curve~\eq{eq:PD:VTPGc}
and the data for higher values of $G$. The origin of this discrepancy
is presumably the same as in the CP. Note also that only four
numerical values are presented for $r_{0}=0.25$: for $G\gtrsim3t$ the
segregated configuration is realized for all $n<1$. This is not
unexpected, as in this case we have the smallest $n^{c}_0$, and thus
largest shifting of the $f$-level, for given $n$. 

As illustrated in \fig{fig:CP2}, the division of electrons between the
two orbitals in the CP strongly affects the position of the critical
line for segregation: the more $f$ electrons relative to
$c$ electrons, the smaller the value of $G$ required to cause
segregation. For $n^{c}_{0}<\half$, turning on the interaction in the
VTP will shift the $f$-level to a lower energy relative to the
$c$-electron band, thus causing a transfer of electrons from the $c$
to the $f$ orbitals. Accordingly, we find that
segregation in the VTP occurs at a lower value of $G$ than in the
$r=r_{0}$ CP [\fig{fig:VTP2}]. 
Eventually, the $f$-level will be shifted below the bottom of the
$c$-electron band; this is the case for couplings
\beq
G_{\text{\tiny{full}}} \geq \frac{-4t[1-\cos(r_{0}n\pi)]}{2n^{c}_{0}-1}
\eeq
The absence of any $c$ electrons to cause 
crystallization or segregation means that any $f$-electron
configuration is the ground state. For $r_{0}=0.25$  and $n=1$ the
critical coupling $G_{\text{\tiny{full}}}\approx2.34t$, explaining the
absence of any $G>2t$ data.

Conversely, for $\epsilon_{f}>0$ and hence $n^{c}_{0}>\half$, turning
on the interaction will shift the $f$-level to higher energies and
empty the $f$-electron orbitals. Segregation may not occur in this case
at all, and the large-$G$ configuration is the empty phase. From our
analysis, we estimate that this will be realized for
\beq
G_{\text{\tiny{empty}}} \geq \frac{-4t[\cos(n\pi)-\cos(r_{0}n\pi)]}{2n^{c}_{0}-1}
\eeq
Note that we assume $n>\half$. 
This scenario is strongly supported by Farka\v{s}ovsk\'{y}'s
numerical study of the $n=1$ VTP.~\cite{F95a} In his
$\epsilon_{f}$-$G$ phase diagram, he found that for $\epsilon_{f}>0$
($n^{c}_{0}>\half$) all electrons occupy the $c$-orbital states for
sufficiently large $G$, while for $\epsilon_{f}<0$ ($n^{c}_{0}<\half$)
the $f$ orbitals become fully occupied as $G$ is increased. 

We note in concluding that we have not addressed the case where the
$f$-level does not lie at the Fermi energy in the noninteracting
system. For example, for $\epsilon_{f}=0$ and $n<\half$ the $f$-level
will lie a finite energy above $\eF=-2t\cos(n\pi)$. On the basis of
our analysis, it appears that a non-zero $f$-population will eventually
appear as the $f$-level is shifted in the presence of a finite $G$. 
As $G$ is further increased, the $c$-electron band is
eventually emptied into the $f$-level. Turning on the interactions, we
thus evolve from a state without any $f$ electrons into a state with
all electrons in the $f$ orbitals. We must regard this result with caution: 
since bosonization is an effective field theory for the excitations
about the Fermi energy, it is difficult to include the
localized electrons whenever $\epsilon_{f}\neq\eF$. In particular, the
bosonic wavelength limit $\alpha$ implies an effective bandwidth 
cut-off for the excitations about $\eF$. What happens when
$\epsilon_f$ lies outside of this effective bandwidth is
unclear and we must go beyond the framework of bosonization to
understand this situation.
For this reason, from the point of view of bosonization the VTP is a
more challenging problem than the CP. 


\section{Extensions of the FKM} \label{sec:EFKM}

The FKM is often studied in a modified form with the addition of extra
terms to the basic Hamiltonian~\eq{eq:Intro:FKM}.
The most common extensions are $c$-$f$
hybridization,~\cite{LRP81,Vfinite,Liu&Ho,BG06,BGinprep} $f$-electron 
hopping~\cite{crystaltff,Batista} or the introduction of
spin.~\cite{DMFT,F99,spin} In the first two cases, the extension
has a dramatic effect upon the physics: 
the occupation of each localized orbital is no longer a good quantum
number. The CP and VTP results are applicable only as limiting
behaviour and we cannot easily incorporate these additional terms into
the analysis presented above. As such, below we will briefly consider
several extensions that maintain the ``classical'' nature of the
$f$ electrons: intraorbital nearest-neighbour interactions and the
addition of spin. Our bosonization formalism is very well suited to
assessing the impact of these extensions upon the ground states of the
``bare'' FKM.  

\subsection{Nearest-neighbour interactions} \label{subsec:NNI}

Our study of nearest-neighbour interactions is confined to their
effect upon the CP results. The same conclusions also hold for the VTP so
long as the densities are normal-ordered.

\subsubsection{$c$ electrons} \label{subsubsec:cNNI}

We write the nearest-neighbour interaction between the $c$ electrons
\beq
{\cal{H}}_{cc} = V_{c}\sum_{j}n^{c}_{j}n^{c}_{j+1} \label{eq:EFKM:cNNI}
\eeq
It is sufficient here to examine only the forward-scattering
contributions of this interaction as the Umklapp and backscattering
contributions are only relevant at half-filling~\cite{G:QP1D}. 
Since we are only interested in the
effect of ${\cal{H}}_{cc}$ on the long-wavelength physics of the FKM,
we may apply the continuum-limit approximation and absorb the
interaction into a free-Boson Hamiltonian
\beq
\widetilde{{\cal{H}}}_{0} =
\frac{v{a}}{2\pi}\sum_{j}\left\{\left(\partial_{x}\widetilde{\phi}(x_j)\right)^{2}+\left(\partial_{x}\widetilde{\theta}(x_j)\right)^{2}\right\} \label{eq:EFKM:Ham}
\eeq
The new Bose fields are related to the $V_{c}=0$ fields by the relations
\beqarray
\widetilde{\phi}(x_j) = \frac{1}{\sqrt{K}}\phi(x_j), && \widetilde{\theta}(x_j) 
=  {\sqrt{K}}\theta(x_j) \label{eq:EFKM:fields} \\
v & = & \frac{\vF}{K} \label{eq:EFKM:v}
\eeqarray
where
\beq
K = \frac{1}{\sqrt{1+2V_{c}\vF/(\pi{at^2})}} \label{eq:EFKM:K}
\eeq
The details of this rescaling procedure are identical to the argument
for the forward-scattering sector of the XXZ chain~\cite{LP75}.
Note that for an attractive interaction $V_{c}=-\pi{a}t^2/\vF$ the
velocity of the Boson modes vanishes: this
indicates the break-down of the bosonization method, as the Luttinger
liquid is unstable towards the clustering of the $c$-electrons. We may
expect the SEG phase to be realized whenever this condition
holds. 

The bosonized FKM with the interaction~\eq{eq:EFKM:cNNI}
is identical in appearance to the $V_{c}=0$ bosonized Hamiltonian
[\eq{eq:B:HamFKM}]: the first term in~\eq{eq:B:HamFKM} is however
replaced by~\eq{eq:EFKM:Ham} and the Bose fields in the other terms
are replaced by their scaled forms~\eq{eq:EFKM:fields}.
Our analysis of $\Ham[FKM]+{\cal{H}}_{cc}$ also proceeds in a similar
way to that in~\Sec{sec:CT}, although we rotate the Hilbert space
using a different canonical transform 
\beq
\hat{U} =
\exp\left\{i\frac{Ga\sqrt{K}}{\pi{v}}\sum_{j'}\tau^{z}_{j}\widetilde{\theta}(x_{j'})\right\} \label{eq:EFKM:CT}
\eeq
As before, we find the effective segregating
interaction~\eq{eq:CT:EffInt}, but with the coefficient changed by
a multiplicative factor
\beq
\frac{G^2a^2}{2\pi\vF}\rightarrow\frac{G^2a^2}{2\pi\vF}\frac{1}{1+2V_{c}a/\pi\vF}
\label{eq:EFKM:EffInt}
\eeq
As we can see, the effect of a repulsive (attractive)
nearest-neighbour interaction between the $c$ electrons is to suppress
(enhance) the segregating interaction. This conclusion is not
surprising: the interaction~\eq{eq:EFKM:cNNI} rescales the charge
compressibility $\kappa$ of the $c$ electrons
$$
\kappa = \kappa_{0}\frac{\vF{K}}{v} = \kappa_{0}K^{2}
$$
where $\kappa_{0}$ is the compressibility for $V_{c}=0$. 
Repulsive interactions ($K<1$) reduce the compressibility and
\emph{vice versa}. In the SEG phase the density of the $c$ electrons
is enhanced due to their confinement to a fraction $1-n^{f}$ of
the lattice. As such, a reduced (enhanced) $c$-electron
compressibility will resist (assist) the formation of this state.

We can easily judge the effect of ${\cal{H}}_{cc}$ on the position of
the critical line $G_{c}$. Following the same arguments as
in~\Sec{subsubsec:seg}, we find in the limit $n^{c}\rightarrow{0}$ the
asymptotic form  
\beq
G_{c} = \left(\frac{\vF}{a}+\frac{2V_{c}}{\pi}\right)\sqrt{\frac{2\pi{A}}{\alpha\delta_{\alpha}(a)}}
\eeq
The term under the square-root is constant for small $c$-electron
fillings; the expression in brackets therefore determines the
small-$n^{c}$ form of the critical line. 
We thus find that for $V_{c}>0$, the SEG phase is only realized above
a finite Coulomb repulsion even in the limit of vanishing $c$-electron
concentration. For attractive interactions, however, the system is
unstable towards segregation for any $c$-electron filling such that
$K^{-1}=0$.


\subsubsection{$f$ electrons} \label{subsubsec:fNNI}

The nearest-neighbour interaction between the $f$ electrons is much
easier to analyze. It is a simple matter to write the interaction term
in the pseudospin representation
\beq
{\cal{H}}_{ff} = V_{f}\sum_{j}n^{f}_{j}n^{f}_{j+1}
= V_{f}\sum_{j}\tau^{z}_{j}\tau^{z}_{j+1} \label{eq:EFKM:fNNI}
\eeq
This nearest-neighbour Ising interaction may be immediately
incorporated into our effective pseudospin model~\eq{eq:EH:EffHam}. 

Quite clearly, an attractive interaction potential $V_{f}<0$ will
make the system unstable towards the SEG phase even for
$G=0$. Crystallization may still occur, although only at finite
coupling strength. Furthermore, we expect that the SEG phase will be
realized even at half-filling for sufficiently large $G$. We consider a
large-$G$ expansion where we project the FKM into a truncated
basis excluding simultaneous occupation of both the $c$ and
$f$ orbitals. To first order in $G^{-1}$ we find the effective
strong-coupling Hamiltonian
\beq
\Ham[SC] = \left(\frac{2t^2}{G}+V_{f}\right)\sum_{j}\widetilde{\tau}^{z}_{j}\widetilde{\tau}^{z}_{j+1}
\eeq
where $\widetilde{\tau}^{z}_{j}=\half(n^{c}_{j}-n^{f}_{j})$ and the
magnetization is fixed at $m^{z}=\half(n^{c}-n^{f})$. The sign of the
nearest-neighbour interaction is ferromagnetic for $G/t>-2t/V_{f}$
implying the formation of the SEG phase. Of course, higher-order [at
least ${\cal{O}}(G^{-2})$] terms complicate this analysis, but by
increasing $G$ we can make their contribution arbitrarily
small.

More interesting is the case of a repulsive potential
$V_{f}>0$. Here~\eq{eq:EFKM:fNNI} hinders segregation, and for 
sufficiently strong $V_{f}$ may suppress it entirely. This is
dependent upon the sign of the nearest-neighbour Ising
interaction in the $\Ham[FKM]+{\cal{H}}_{ff}$ effective pseudospin
model: 
the SEG phase cannot be realized unless the nearest-neighbour Ising
interaction is ferromagnetic. By equating~\eq{eq:PD:nnapprox}
and~\eq{eq:EFKM:fNNI} we immediately find a condition for
the appearance of segregation:
\beq
G>\sqrt{\frac{2\pi\vF{V_f}}{a^2\delta_{\alpha}(a)}} \label{eq:EFKM:ferro}
\eeq
Since for $n^{c}\ll{1}$ we have
$\delta_{\alpha}(a)\sim\alpha^{-1}\gtrsim{\cal{O}}(\kF)$, the RHS
of~\eq{eq:EFKM:ferro} should be finite at low $c$-electron filling. 
Phase separation as discussed in~\Sec{subsubsec:PS} will nevertheless
still occur as the range of the interaction~\eq{eq:CT:EffInt} extends
beyond nearest-neighbours, with these higher-order terms in the
pseudospin model remaining unaffected by the addition of
${\cal{H}}_{ff}$. The most important of these extra terms is the
next-nearest-neighbour interaction:
if the
condition~\eq{eq:EFKM:ferro} is not satisfied, this term is the
dominant ferromagnetic coupling, and hence orders
the $f$-electrons into a single cluster where only every \emph{second}
site is occupied. That is, we may expect that a large
portion of the SEG phase in the phase diagram~\fig{fig:CP2} will be
replaced by a phase-separation between the empty and period-$2$
crystalline configurations. 

Numerical results for the FKM with nearest-neighbour $f$-electron
repulsion confirm this scenario: Gajek and Lema\'{n}ski have studied 
the effect of~\eq{eq:EFKM:fNNI} in the canonical ensemble for
$V_{f}=0.1G$.~\cite{GL03} For a repulsive potential of this form, the
SEG phase was not realized at {\it{any}} coupling strength or electron
filling. For $n^{c}\ll{1}$, the $f$-electrons indeed phase
separate into the period-$2$ crystalline and empty configurations. 
Interestingly, for the $n^{c}_{0}=n^{f}_{0}$ case presented, phase separation
is realized only for $n<0.4$. This indicates is a significant truncation of
the range of the segregating interaction with increasing filling. 

\subsection{Spin} \label{subsec:Spin}

To use the FKM as a model of any realistic condensed-matter system, we
are required to relax the assumption of spinless
electrons. Simply adding a spin index to the fermionic operators 
in~\eq{eq:Intro:FKM} is, however, not enough: we must take into
account the orbital structure of the localized states. Because of
the small radius of the $f$ orbitals, the intra-ionic correlations are
very strong, prompting us to introduce a Coulomb repulsion $U$ between
the $f$ electrons in our spinful model. We thus write
\beq
\Ham =
-t\sum_{j}\sum_{\sigma}\left\{c^{\dagger}_{j,\sigma}c^{}_{j+1,\sigma} +\Hc\right\}
+ U\sum_{j}n^{f}_{j,\uparrow}n^{f}_{j,\downarrow} 
+ G\sum_{j}\sum_{\sigma,\sigma'}n^{f}_{j,\sigma}n^{c}_{j,\sigma'}
\label{eq:EFKM:spinFKM} 
\eeq
We consider here only the limit $U=\infty$ where double
occupation of an $f$-orbital is excluded from the physical
subspace. This situation has been numerically studied by
Farka\v{s}ovsk\'{y} for the fixed total electron concentration $n=1$
in both the CP and VTP interpretations.~\cite{F99} Since
double occupation is forbidden, we may represent the $f$ operators in
terms of spinless fermion (holon) operators $e_{j}$:~\cite{tJ} 
\beq
\sum_{\sigma}n^{f}_{j,\sigma} = (1-e^{\dagger}_{j}e^{}_{j})
\eeq
That is, at any site without an $f$-electron we find a spinless hole.
We hence re-write~\eq{eq:EFKM:spinFKM}
\beq
\Ham =
-t\sum_{j}\sum_{\sigma}\left\{c^{\dagger}_{j,\sigma}c^{}_{j+1,\sigma}+\Hc\right\}
+ G\sum_{j}\sum_{\sigma}(1-e^{\dagger}_{j}e^{}_{j})n^{c}_{j,\sigma} \label{eq:EFKM:U=infty}
\eeq
The $f$-orbital occupation is thus described by spinless fermions as
in the usual FKM; the condition for fixed total electron concentration
is however written
$n=(1/N)\sum_{j}[1-\langle{e^{\dagger}_{j}e^{}_{j}}\rangle+\sum_{\sigma}\langle{n^{c}_{j,\sigma}}\rangle]$. 
The spin-modes of the $c$ electrons cannot be removed as for the
$f$ electrons.

The bosonization procedure outlined in~\Sec{sec:B} requires little
modification to include the spin degrees of freedom. We define boson
fields in terms of the density fluctuations $\rho_{\nu,\sigma}(k)$ 
in each spin-channel
\beqarray
\phi_{\sigma}(x_j) & = &
-i\sum_{\nu}\sum_{k\neq{0}}\frac{\pi}{kL}\rho_{\nu,\sigma}(k)\CO{}e^{-ikx_j}
\\
\theta_{\sigma}(x_j) & = &
i\sum_{\nu}\sum_{k\neq0}\nu\frac{\pi}{kL}\rho_{\nu,\sigma}(k)\CO{}e^{-ikx_j} 
\eeqarray
Boson fields with different spin-indices commute; fields with the same
spin-indices obey the commutation relations~\eq{eq:B:comphitheta}
and~\eq{eq:B:comdelphitheta}. 
It is convenient to split the bosonic representation into
charge- and spin-sectors, defined respectively by the linear
combinations 
\beqarray
\phi_{c}(x_j) & = &
\frac{1}{\sqrt{2}}[\phi_{\uparrow}(x_j)+\phi_{\downarrow}(x_j)] \\ 
\phi_{s}(x_j) & = &
\frac{1}{\sqrt{2}}[\phi_{\uparrow}(x_j)-\phi_{\downarrow}(x_j)] 
\eeqarray
and similarly for the $\theta$-fields. This will considerable simplify
the bosonic representation of our Hamiltonian.
After some algebra we find the bosonic representation for the electron
density operator
\beq
\sum_{\sigma}n^{c}_{j,\sigma} = n^{c}_{0}
-\frac{\sqrt{2}a}{\pi}\partial_{x}\phi_{c}(x_j) +
\frac{4Aa}{\alpha}\cos[\sqrt{2}\phi_{s}(x_j)]\cos[\sqrt{2}\phi_{c}(x_j)-2\kF{x_j}]
\eeq
where $\kF=\pi{n^{c}}/2a$.
Note that the forward-scattering contribution (second term on RHS) is
very similar as in 
the spinless case~\eq{eq:B:b_nc}; the backscattering contribution
(third term on RHS) however involves both the spin- and charge-sector
fields. 

Again we adopt a pseudospin representation for the $f$-orbital
occupation: we define $\tau^{z}_{j} = \half-e^{\dagger}_{j}e^{}_{j}$, and
so as before spin-$\uparrow$ corresponds to an occupied orbital and
spin-$\downarrow$ to an empty site. Following the same basic
manipulations as for the spinless case, we obtain the bosonized
Hamiltonian
\beqarray
\Ham &=
&\frac{\vF{a}}{2\pi}\sum_{\xi=c,s}\sum_{j}\left\{\left(\partial_{x}\phi_{\xi}(x_j)\right)^{2}+\left(\partial_{x}\theta_{\xi}(x_j)\right)^{2}\right\}+G(n^{c}_{0}-\half)\sum_{j}\tau^{z}_{j} 
\notag \\ 
&&-\frac{\sqrt{2}Ga}{\pi}\sum_{j}\tau^{z}_{j}\partial_{x}\phi_{c}(x_j)
\notag\\
&&+\frac{4GAa}{\alpha}\sum_{j}\tau^{z}_{j}\cos[\sqrt{2}\phi_{s}(x_j)]\cos[\sqrt{2}\phi_{c}(x_j)-2\kF{x_j}] \label{eq:EFKM:BspinFKM}
\eeqarray
Excluding the last term,~\eq{eq:EFKM:BspinFKM} is identical
to its spinless equivalent. Importantly, the
forward-scattering interaction (second last term) remains in the
same form as before: by simply shifting the charge-sector boson
frequencies we may remove this term. This requires us to apply the
canonical transform 
\beq
\hat{U} =
\exp\left\{i\frac{\sqrt{2}Ga}{\pi\vF}\sum_{j'}\tau^{z}_{j'}\theta_{c}(x_{j'})\right\}  
\eeq
After some algebra, we obtain the transformed Hamiltonian
\beqarray
\hat{U}^{\dagger}\Ham\hat{U} & = &
\frac{\vF{a}}{2\pi}\sum_{\xi=c,s}\sum_{j}\left\{\left(\partial_{x}\phi_{\xi}(x_j)\right)^{2}+\left(\partial_{x}\theta_{\xi}(x_j)\right)^{2}\right\}+G(n^{c}_{0}-\half)\sum_{j}\tau^{z}_{j}
\notag \\
& &
-\frac{G^2a^2}{\pi\vF}\sum_{j,j'}\tau^{z}_{j}\delta_{\alpha}(x_j-x_{j'})\tau^{z}_{j'}
\notag \\
&&
+\frac{4GAa}{\alpha}\sum_{j}\tau^{z}_{j}\cos[\sqrt{2}\phi_{s}(x_j)]\cos[\sqrt{2}\phi_{c}(x_j)-2{\cal{K}}(j)-2\kF{x_j}] \label{eq:EFKM:CTHam}
\eeqarray
where ${\cal{K}}(j)$ is defined as in~\eq{eq:CT:calK}. Since the
$c$-electron spin modes do not contribute to the physics, we may
replace $\phi_{s}$ by its noninteracting expectation value, i.e
$\phi_{s}=\langle{\phi_s}\rangle=0$. Substituting this
into~\eq{eq:EFKM:BspinFKM} we obtain the same 
effective Hamiltonian as for the spinless FKM. This
allows us to draw an important conclusion: for the spinful
model~\eq{eq:EFKM:U=infty} with $f$- and $c$-electron concentrations
$n^{f}$ and $n^{c}$ respectively, the configuration adopted by the
$f$ electrons is \emph{identical} to that adopted by the spinless
$f$ electrons in~\eq{eq:Intro:FKM} with $f$- and $c$-electron
concentrations $n^{f}$ and $\half{n^{c}}$ respectively. This explains
the appearance of phase separation and segregation in the numerical
study of~\eq{eq:EFKM:U=infty} at $n^{f}+n^{c}=1$.~\cite{F99}

The obvious extension of~\eq{eq:EFKM:spinFKM} would be the
inclusion of a Kondo-like exchange between the $c$ and $f$ electrons
on each site $j$. This model may be useful for understanding the
properties of the manganites, which are known to display a phase
separated state.~\cite{KRS05} Such a model could display an interesting
coexistence of spin- and charge-order; this problem remains to be fully
addressed.~\cite{spin,GMcCJR04} 

\section{Conclusions} \label{sec:conclusions}

In this paper we have presented a novel approach to the study of
charge order in the FKM below half-filling. We used a bosonization
method that accounted for the non-bosonic density fluctuations of the
$c$ electrons below a certain  
length-scale $\alpha>a$; we identified $\alpha$ as characterizing the 
delocalization of the $c$ electrons. This
delocalization of the $c$ electrons over several lattice sites favours
empty underlying $f$ orbitals in order to minimize the interorbital Coulomb
repulsion. We demonstrated in~\Sec{subsec:SEGint} how this directly
leads to effective attractive interactions between the $f$ electrons,
and hence the SEG phase.
Using a canonical transform, we obtained an explicit form for the
segregating interaction~\eq{eq:CT:EffInt}.
Since the canonical transform was carried out to infinite order, this
interaction is non-perturbative. 
The canonical transform is a generalization of the transform used in
Schotte and Schotte's solution of the XEP.~\cite{SS69} 
We argued in~\Sec{subsubsec:KLM} for a parallel
between~\eq{eq:CT:EffInt} and the double-exchange interaction in the
KLM, based upon the similar orthogonality catastrophe physics in the
single-impurity limit of both models.  

The canonical transform permitted a decoupling of the $c$ and
$f$ electrons, yielding 
an effective Ising model [\eq{eq:EH:EffHam}] for the configuration of
the $f$ electrons, based upon a pseudospin-$\half$ representation for 
the occupation of the localized orbitals. This effective model,
$\Ham[eff]$, 
clearly revealed the competition between the backscattering
crystallization and the forward-scattering segregation. 
$\Ham[eff]$ correctly predicted the structure of the CP phase
diagram: we obtained an expression for the  
critical coupling required for segregation which is in good agreement
with the numerical results. 
We also demonstrated that the effective model could successfully
account for the instability towards phase separation between a
crystalline and the empty phase in the weak-coupling FKM. Our approach
was not limited to the CP, and in~\Sec{subsec:VTP} we considered the
phase diagram of the VTP. We found that the Coulomb repulsion shifted
the bare $f$-level, causing a ``classical'' valence transition. The
sign of the $f$-level shift is highly dependent upon the
band structure. Finally, we discussed the impact of intraorbital
nearest-neighbour interactions (\Sec{subsec:NNI}) and the
introduction of spin (\Sec{subsec:Spin}) on the charge order found in
the spinless model. 

Prospects for future work are promising. 
We have already outlined an application of our method to the nontrivial
extension of~\eq{eq:Intro:FKM} by the addition of an on-site
hybridization potential, the so-called Quantum FKM (QFKM).~\cite{BG06} 
The crystallization is heavily suppressed in the QFKM, as the dominant
feature of the $c$-electron behaviour at weak-coupling is the resonant
scattering off the $f$ orbitals (mixed-valence). 
In contrast, segregation occurs in the QFKM, as the responsible
orthogonality catastrophe physics remains 
intact with the introduction of the hybridization 
(\Sec{subsubsec:KLM}). Here, however, we expect dynamic
charge-screening processes (in analogy to the spin-screening in the
KLM), with important and interesting consequences that we will fully explore
in a forthcoming publication.~\cite{BGinprep}

\section*{Acknowledgments}

P.M.R.B. thanks C. D. Batista, A. R. Bishop and J. E.Gubernatis for
useful discussions. The authors thank M. Bortz for his critical
reading of the manuscript.

\end{document}